\documentclass[twocolumn]{aastex701}

\usepackage{makecell, mathtools}
\usepackage{hyperref}
\usepackage{multirow}

\begin{document}

\title{First Pre-peak Ultraviolet Spectrum of a Tidal Disruption Event:\\A Fast Outflow Revealed Prior to Maximum Light in TDE\,2025aarm}
\shorttitle{TDE\,2025aarm from Near-Infrared to X-rays}

\author[0000-0002-5698-8703]{Erica Hammerstein}
\affiliation{Department of Astronomy, University of California, Berkeley, CA 94720-3411, USA}
\affiliation{Berkeley Center for Multi-messenger Research on Astrophysical Transients and Outreach (Multi-RAPTOR), University of California, Berkeley, CA 94720-3411, USA}
\email[show]{ekhammer@berkeley.edu}

\author[0000-0002-7706-5668]{Ryan Chornock}
\affiliation{Department of Astronomy, University of California, Berkeley, CA 94720-3411, USA}
\affiliation{Berkeley Center for Multi-messenger Research on Astrophysical Transients and Outreach (Multi-RAPTOR), University of California, Berkeley, CA 94720-3411, USA}
\email{chornock@berkeley.edu}

\author[0000-0001-8426-5732]{Jean Somalwar}
\affiliation{Department of Astronomy, University of California, Berkeley, CA 94720-3411, USA}
\affiliation{Berkeley Center for Multi-messenger Research on Astrophysical Transients and Outreach (Multi-RAPTOR), University of California, Berkeley, CA 94720-3411, USA}
\affiliation{Kavli Institute for Particle Astrophysics and Cosmology, Stanford, CA 94305, USA}
\email{jsomalwar@berkeley.edu}

\author[0000-0003-4768-7586]{Raffaella Margutti}
\affiliation{Department of Astronomy, University of California, Berkeley, CA 94720-3411, USA}
\affiliation{Berkeley Center for Multi-messenger Research on Astrophysical Transients and Outreach (Multi-RAPTOR), University of California, Berkeley, CA 94720-3411, USA}
\affiliation{Department of Physics, University of California, 366 Physics North MC 7300, Berkeley, CA 94720, USA}
\email{rmargutti@berkeley.edu}

\author[0000-0001-6350-8168]{Brenna Mockler}
\affiliation{Department of Physics \& Astronomy, University of California, Davis, CA 95616, USA}
\email{bmockler@ucdavis.edu}

\author[0000-0001-5674-8403]{Olivia Aspegren}
\affiliation{Department of Astronomy, University of California, Berkeley, CA 94720-3411, USA}
\email{oliviaaspegren@berkeley.edu}

\author[0000-0003-0466-3779]{Itai Sfaradi}
\affiliation{Department of Astronomy, University of California, Berkeley, CA 94720-3411, USA}
\affiliation{Berkeley Center for Multi-messenger Research on Astrophysical Transients and Outreach (Multi-RAPTOR), University of California, Berkeley, CA 94720-3411, USA}
\email{itai.sfaradi@berkeley.edu}

\author[0000-0002-2249-0595]{Natalie LeBaron}
\affiliation{Department of Astronomy, University of California, Berkeley, CA 94720-3411, USA}
\affiliation{Berkeley Center for Multi-messenger Research on Astrophysical Transients and Outreach (Multi-RAPTOR), University of California, Berkeley, CA 94720-3411, USA}
\email{nlebaron@berkeley.edu}

\author[0000-0002-8977-1498]{Jonathan Carney}
\affiliation{Department of Physics and Astronomy, University of North Carolina at Chapel Hill, Chapel Hill, NC 27599-3255, USA}
\email{jcarney@unc.edu}

\author[0000-0003-3460-0103]{Alexei V. Filippenko}
\affiliation{Department of Astronomy, University of California, Berkeley, CA 94720-3411, USA}
\email{afilippenko@berkeley.edu}

\author[0000-0002-8070-5400]{Nayana AJ}
\affiliation{Department of Astronomy, University of California, Berkeley, CA 94720-3411, USA}
\affiliation{Berkeley Center for Multi-messenger Research on Astrophysical Transients and Outreach (Multi-RAPTOR), University of California, Berkeley, CA 94720-3411, USA}
\email{nayana@berkeley.edu}

\author[0000-0002-8977-1498]{Igor Andreoni}
\affiliation{Department of Physics and Astronomy, University of North Carolina at Chapel Hill, Chapel Hill, NC 27599-3255, USA}
\email{igor.andreoni@unc.edu}

\author[0000-0001-5955-2502]{Thomas Brink}
\affiliation{Department of Astronomy, University of California, Berkeley, CA 94720-3411, USA}
\email{tgbrink@berkeley.edu}

\author[0000-0001-6272-5507]{Peter J. Brown}
\affiliation{Department of Physics and Astronomy, Texas A\&M University, 4242 TAMU, College Station, TX 77843, USA}
\affiliation{George P. and Cynthia Woods Mitchell Institute for Fundamental Physics \& Astronomy, Texas A\&M University, 4242 TAMU, College Station, TX 77843, USA}
\email{pbrown801@tamu.edu}

\author[0000-0002-6523-9536]{Adam J.~Burgasser}
\affiliation{Department of Astronomy \& Astrophysics, UC San Diego, La Jolla, CA 92093, USA}
\email{aburgasser@ucsd.edu}

\author[0000-0003-1673-970X]{S. Bradley Cenko}
\affiliation{Joint Space-Science Institute, University of Maryland, College Park, MD 20742, USA}
\affiliation{Astrophysics Science Division, NASA Goddard Space Flight Center, Mail Code 661, Greenbelt, MD 20771, USA}
\email{brad.cenko@nasa.gov}

\author[0000-0003-3703-5154]{Suvi Gezari}
\affiliation{Department of Astronomy, University of Maryland, College Park, MD 20742, USA}
\email{suvi@umd.edu}

\author[0000-0003-2868-489X]{Xiaoshan Huang}
\affiliation{California Institute of Technology, TAPIR, Mail Code 350-17, Pasadena, CA 91125, USA}
\email{xshuang@caltech.edu}

\author[0000-0002-5619-4938]{Mansi M. Kasliwal}
\affil{Division of Physics, Mathematics, and Astronomy, California Institute of Technology, Pasadena, CA 91125, USA}
\email{mansi@astro.caltech.edu}

\author[0000-0003-2869-7682]{Jon M. Miller}
\affiliation{Department of Astronomy, University of Michigan, 1085 South University Avenue, Ann Arbor, MI 48109, USA}
\email{jonmm@umich.edu}

\author[0009-0009-0147-6485]{Sasha Mintz}
\affiliation{Department of Physics \& Astronomy, University of Southern
California, Los Angeles, California, 90007, USA}
\affiliation{Observatories of the Carnegie Institution for Science, 813 Santa Barbara St., Pasadena, CA 91101, USA}
\email{smintz@carnegiescience.edu}

\author[0000-0003-4725-4481]{Sam Rose} 
\affiliation{Division of Physics, Mathematics, and Astronomy, California Institute of Technology, Pasadena, CA 91125, USA}
\email{srose@caltech.edu}

\author[0000-0002-9132-6561]{Peter Senchyna}
\affiliation{Observatories of the Carnegie Institution for Science, 813 Santa Barbara St., Pasadena, CA 91101, USA}
\email{psenchyna@carnegiescience.edu}

\author[0000-0002-4733-4994]{Joshua D. Simon}
\affiliation{Observatories of the Carnegie Institution for Science, 813 Santa Barbara St., Pasadena, CA 91101, USA}
\email{jsimon@carnegiescience.edu}

\author[0000-0002-1420-1837]{Emma Softich}
\affiliation{Department of Astronomy \& Astrophysics, UC San Diego, La Jolla, CA 92093, USA}
\email{esoftich@ucsd.edu}

\author[0000-0003-2434-0387]{Robert Stein}
\affiliation{Department of Astronomy, University of Maryland, College Park, MD 20742, USA}
\affiliation{Joint Space-Science Institute, University of Maryland, College Park, MD 20742, USA}
\affiliation{Astrophysics Science Division, NASA Goddard Space Flight Center, Mail Code 661, Greenbelt, MD 20771, USA}
\email{rdstein@umd.edu}

\author[0000-0001-6747-8509]{Yuhan Yao}
\affiliation{Department of Astronomy, University of California, Berkeley, CA 94720-3411, USA}
\affiliation{Berkeley Center for Multi-messenger Research on Astrophysical Transients and Outreach (Multi-RAPTOR), University of California, Berkeley, CA 94720-3411, USA}
\affiliation{Miller Institute for Basic Research in Science, 206B Stanley Hall, Berkeley, CA 94720, USA}
\email{yyao@pku.edu.cn}

\author[0000-0002-2636-6508]{WeiKang Zheng}
\affiliation{Department of Astronomy, University of California, Berkeley, CA 94720-3411, USA}
\email{weikang@berkeley.edu}

\author[0000-0002-5884-7867]{Richard Dekany}
\affiliation{Caltech Optical Observatories, California Institute of Technology, Pasadena, CA  91125}
\email{rgd@astro.caltech.edu}

\author[0000-0002-3168-0139]{Matthew J. Graham}
\affiliation{Cahill Center for Astrophysics, California Institute of Technology, 1216 East California Boulevard, Pasadena, CA 91125, USA}
\email{mjg@caltech.edu}

\author[0000-0003-2451-5482]{Russ R. Laher}
\affiliation{IPAC, California Institute of Technology, 1200 E. California Blvd, Pasadena, CA 91125, USA}
\email{laher@ipac.caltech.edu}

\author[0000-0003-1227-3738]{Josiah Purdum}
\affiliation{Caltech Optical Observatories, California Institute of Technology, Pasadena, CA 91125}
\email{jpurdum@caltech.edu}

\author[0000-0001-7648-4142]{Ben Rusholme}
\affiliation{IPAC, California Institute of Technology, 1200 E. California
             Blvd, Pasadena, CA 91125, USA}
\email{rusholme@ipac.caltech.edu}

\author[0000-0003-1546-6615]{Jesper Sollerman}
\affiliation{The Oskar Klein Centre, Department of Astronomy, Stockholm University, AlbaNova, SE-10691 Stockholm, Sweden}
\email{jesper@astro.su.se}

\author[0000-0001-9276-1891]{St\'efan van der Walt}
\affiliation{Berkeley Institute for Data Science, University of California Berkeley, Berkeley, CA 94720, USA}
\email{stefanv@berkeley.edu}

\begin{abstract}
The tidal disruption and eventual accretion of a star by a massive black hole can lead to the launching of outflows and winds that encode important information about the tidal disruption and accretion processes. However, little is known about the existence and behavior of these outflows before optical light-curve peak, when processes such as stream-stream collisions and disk formation may be important.
Here, we present the first pre-peak ultraviolet (UV) spectrum of a tidal disruption event (TDE), obtained $\sim 20$ days before optical maximum light of the nearby (redshift $z=0.0137$) TDE\,2025aarm, along with quasi-simultaneous infrared (IR) through X-ray observations.
Our HST/STIS spectrum shows strong evidence for an early-time outflow through broad near-UV (NUV) and far-UV (FUV) absorption lines blueshifted by $\sim$10,000 km s$^{-1}$, including two new UV broad absorption features not yet identified in a TDE. We find that the FUV--IR continuum deviates significantly from a blackbody ($f_\lambda \propto \lambda^{-3.08}$), which we interpret as a signature of reprocessing through the outflow. This deviation implies that the bolometric luminosity in optical TDEs may be underestimated by a significant factor ($\sim$9 in this case) when inferred from single-temperature blackbody fits to NUV--optical photometry alone.
This work further confirms that TDEs are capable of launching fast outflows at very early times and emphasizes the importance of prompt FUV spectroscopic observations of TDEs that can capture the full continuum emission and energetics.

\end{abstract}

\keywords{\uat{Accretion}{14} --- \uat{Supermassive black holes}{1663} --- \uat{Black hole physics}{159} --- \uat{Ultraviolet spectroscopy}{2284}}

\section{Introduction}

The tidal disruption of a star by a supermassive black hole \citep[SMBH; e.g.,][]{Hills75, Rees88} can produce a luminous flare of radiation detectable across the electromagnetic spectrum. While early theoretical work predicted that these tidal disruption events (TDEs) would primarily emit as X-ray transients \citep{Rees88}, many TDEs are now discovered at optical wavelengths thanks to wide-field surveys such as ASAS-SN \citep{ASASSN}, ATLAS \citep{Tonry2018}, PanSTARRS \citep{chambers16}, and the Zwicky Transient Facility \citep[ZTF;][]{bellm19, Graham19, Dekany20, masci19}. These multi-wavelength samples of TDEs have revealed a dichotomy in their X-ray and optical/ultraviolet (UV) properties. The X-ray properties are largely consistent with thermal emission from the nascent accretion disk, while the optical emission is dominated by a $\sim10^4-10^5$ K blackbody continuum which arises from much larger radii than would be expected for the newly formed disk (e.g., \citealt{vanVelzen21}, \citealt{Hammerstein23}, or for a review, \citealt{Gezari2021}).

Several possible origins for the optical/UV emission have arisen, including shocks and outflows created by intersecting stellar debris streams \citep[e.g.,][]{lu20, Piran2015, jiang16} and the reprocessing of accretion-disk emission by an outflow or disk wind launched after material begins accreting onto the black hole \citep[e.g.,][]{strubbe09, metzger16, Dai2018}. In the latter scenario, the optical emission is dominated by a heated, optically thick outflow driven away from a super-Eddington accretion disk \citep{Bu2022}. Indeed, follow-up observations of TDEs have revealed winds and outflows across a wide range of velocities. In the event ASASSN-14li \citep{Holoien16_14li}, for example, outflows were observed at X-ray and UV wavelengths at velocities of a few hundred km s$^{-1}$ to 0.2$c$ \citep{Miller2015, kara18, cenko16}. The UV spectrum of ASASSN-14li is characterized largely by broad emission features. Other events, such as iPTF\,15af and AT\,2019qiz \citep{Blagorodnova2019, Hung2021}, are instead characterized by strong, blueshifted, broad absorption lines (BALs) that are consistent with $\approx$5000--15,000 km s$^{-1}$ outflow velocities. BALs are detected in many TDEs with UV spectra \citep{Chornock2014, brown18, Blagorodnova2019, Hung2021}, but even among these there is a diversity in the ionization species, blueshift velocities, and line widths detected. For example, iPTF\,15af exhibited FUV BALs of \ion{Si}{4}, \ion{C}{4}, \ion{N}{5}, and tentative \ion{P}{5} blueshifted by $\lesssim6000$ km s$^{-1}$ \citep{Blagorodnova2019}. In contrast, AT\,2019qiz showed additional NUV BALs of Fe at early times (within a few weeks after light curve peak) which subsided by 65 days after light-curve peak \citep{Hung2021}. 

The diversity in TDE UV spectra, primarily regarding whether emission or absorption lines are observed, has been explained as a viewing-angle effect, where sight lines through the wind produce BALs, and sight lines above or below the wind produce emission features \citep{Parkinson2020}. The evolution of BALs in TDE UV spectra, as in AT\,2019qiz from 13 days to 98 days after light-curve peak, has been interpreted as the thinning of the X-ray/extreme-UV (EUV) shielding material as the mass fallback and accretion rates decrease with time \citep{Hung2021}. This evolution is rich with information on the debris-stream interactions, outflow behavior, and accretion in TDEs. However, prior to TDE\,2025aarm, there have been no such observations of a UV outflow before maximum light in a TDE.

TDE\,2025aarm/ZTF18acecugr was first reported to the Transient Name Server (TNS) by GOTO \citep{Brutus_GOTO} on 2025 Oct. 17 (UTC), with initial detections on 2025 Sep. 30. TDE\,2025aarm was also detected by ZTF beginning  2025 Sep. 30, but it was not reported to TNS because the initial star/galaxy separation algorithm failed \citep[see][]{ZTF_SGscore}. Additionally, the host galaxy possesses a $\sim$4$\sigma$ Gaia parallax \citep{gaia16,Gaia_DR3}, meaning that even if the source had passed the initial star/galaxy check, it would have been rejected later in the filtering process because it would have been considered a Galactic source\footnote{Changes have been made to the internal alert processing of the ZTF alert stream to avoid these failure modes in the future.}. 

TDE\,2025aarm was classified by \citet{Faris_Brutusclass} on 2025 Oct. 30 as a TDE\,H+He owing to its blue continuum and broad hydrogen and \ion{He}{2} emission features in follow-up optical spectra. Narrow host-galaxy absorption features confirmed a redshift of $z=0.01368$, making TDE\,2025aarm among the nearest TDEs discovered to date at 59 Mpc. Pre-peak follow-up radio observations of TDE\,2025aarm revealed nondetections with upper limits in the 5 and 8 GHz bands of $6.5\times10^{27}$ and $7.8\times10^{27}$ erg s$^{-1}$ Hz$^{-1}$, respectively \citep{Sfaradi_Brutus_radio}. Further radio observations fewer than four days later revealed one of the faintest radio detections of a TDE ever with $1.5\times10^{26}$ erg s$^{-1}$ Hz$^{-1}$ \citep{Brutus_VLA}. At X-ray energies, \citet{Brutus_Chandra} reported detections with the Chandra X-ray Observatory (hereafter Chandra) corresponding to $5.9\times10^{-15}$ erg s$^{-1}$ cm$^{-2}$ in the 0.5--7.0 keV band (observed) more than 20 days pre-optical-peak. This constitutes one of the faintest, earliest X-ray detections of an optically selected TDE yet, which we include as part of this work. Subsequent post-peak Chandra observations ($\gtrsim 40$ days post-peak) indicated that the source brightened significantly in X-rays after optical light-curve peak \citep{Brutus_Chandra2}. In the final stages of preparation before submission of this manuscript, \citet{Aamer2026} released a complementary analysis of the optical spectral and optical--UV SED evolution of TDE\,2025aarm, finding a light curve that deviates from smooth evolution and complex structure in the broad H$\alpha$ line profile.

In this paper, we present near-simultaneous infrared (IR) through X-ray observations of TDE\,2025aarm, including the earliest (relative to optical light-curve peak) UV spectrum of a new TDE obtained to date\footnote{\citet{Payne2023} obtained STIS spectra on the rise of flares 18 and 19 of the repeating nuclear transient ASASSN-14ko. This event has been interpreted as a repeating partial TDE.}. The paper is organized as follows. In Section \ref{sec:data} we present the observations and data reduction. Section \ref{sec:host} discusses the host-galaxy characterization. We detail the data analysis and results in Section \ref{sec:results}, and we end with a discussion of these results in Section \ref{sec:discussion}. Throughout this paper, we adopt a flat cosmological model with $\Omega_\Lambda = 0.7$ and H$_0 = 70$ km s$^{-1}$ Mpc$^{-1}$ \citep{cosmo_cite}. All photometry is presented in the AB system \citep{Oke1983}.

\section{Data} \label{sec:data}

\subsection{Optical/UV Photometry}
TDE\,2025aarm was observed by the Neil Gehrels Swift Observatory \citep{Gehrels2004} in the UV with the Ultra-Violet Optical Telescope \citep[UVOT;][]{Roming2005} over $\sim100$ days from pre- to post-peak (PIs Charalampopoulos, Miller, Sun, and Stein). We reduced these data using standard procedures with a 10\arcsec~aperture for the \texttt{uvotsource} package\footnote{HEASoft 6.34 and UVOT CALDB 20240201}, subtracting the host-galaxy flux estimated from the population synthesis described in Section \ref{sec:host}.

We requested ZTF forced point-spread-function (PSF) photometry for TDE\,2025aarm to extract precise flux measurements of the transient through the ZTF forced-photometry service \citep{Masci_newFP}. Following the recommendations of \citet{Masci_newFP}, we employ quality cuts to filter out epochs affected by bad pixels and apply thresholds for seeing and the sigma-per-pixel in the input science image. We also perform a baseline correction to account for systematic offsets in the forced flux measurements, similar to \citet{Hammerstein23}, finding that all bands require a $\lesssim 26~\mu$Jy shift. We also obtained $ugri$ photometry with the Spectral Energy Distribution Machine \citep[SEDM;][]{Blagorodnova2018, Fremling2016}. In Table \ref{tab:SEDM}, we list available SEDM spectra \citep{Rigault2019} which are not included in this analysis due to their lower resolution but will be made available as part of the data associated with this paper.

\begin{figure}
    \centering
    \includegraphics[width=\linewidth]{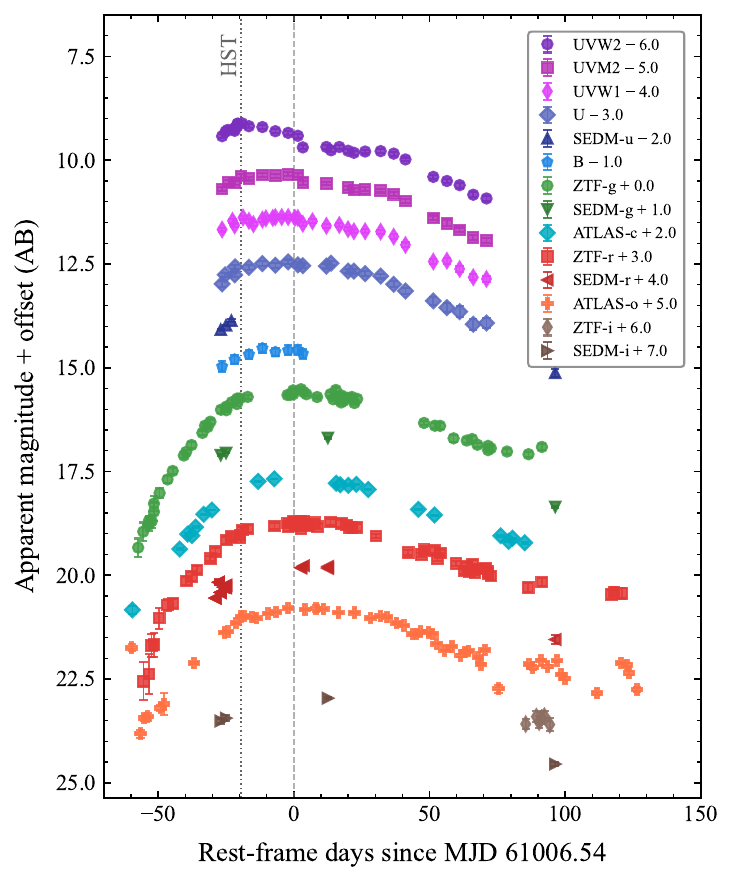}
    \caption{UV--optical light curve of TDE\,2025aarm, including photometry from Swift/UVOT, ZTF, and ATLAS. The transient rises to peak brightness in $\sim60$ days, with a slower decay and the start of a possible plateau at $\Delta t > 60$ days, just before the end of its seasonal visibility window. We mark the time of the pre-peak HST/STIS UV spectrum at $\Delta t \approx -20$ days with the dotted line, the first pre-peak UV spectrum of a TDE ever obtained. The dashed line represents our measured $t_{\rm peak} = \rm MJD~61006.54$.}
    \label{fig:opt-lc}
\end{figure}

We requested additional optical forced photometry of TDE\,2025aarm from the Asteroid Terrestrial-impact Last Alert System (ATLAS) survey using the ATLAS forced-photometry service\footnote{\url{https://fallingstar-data.com/forcedphot/}} \citep{Tonry2018, Smith2020}. For quality control, we remove epochs with significantly negative flux measurements ($< - 100~\mu$Jy) or large uncertainties ($> 15~\mu$Jy), as well as significant outliers. These quality cuts remove 24 data points from the light curve, which do not significantly impact our results. We show the UVOT, ZTF, and ATLAS light curves in Figure \ref{fig:opt-lc}, and describe the light-curve fitting used to obtain $t_{\rm{peak},g}$, $t_{\rm 1/2, rise}$, and $t_{\rm 1/2, decay}$ in Section \ref{SubSec:lcfit}.

\subsection{HST/STIS}
\begin{figure*}
    \centering
    \includegraphics[width=\textwidth]{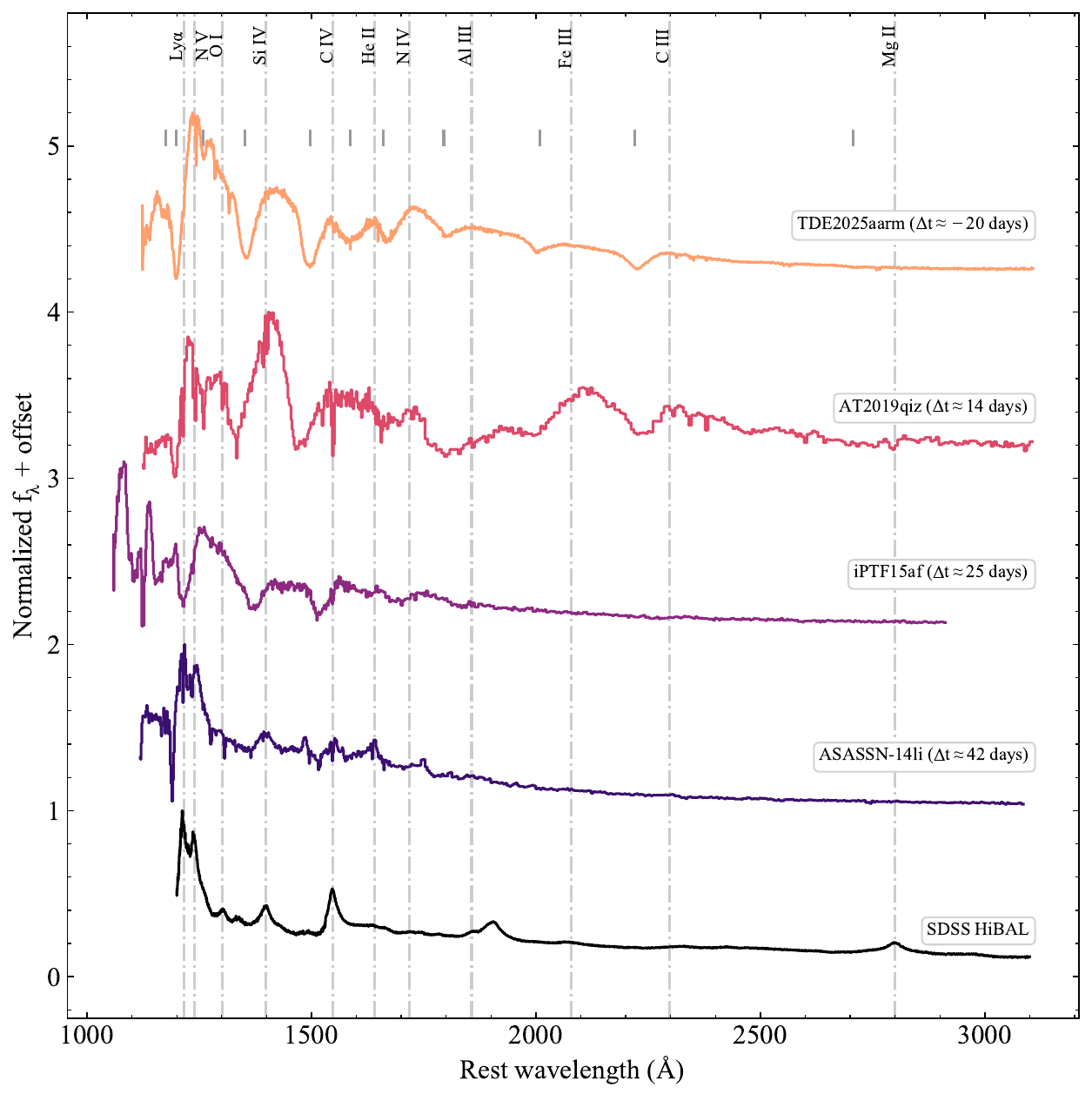}
    \caption{UV spectra of selected TDEs with BALs or emission lines, labeled with $\Delta t = t_{\rm spec} - t_{\rm peak}$. The single gray dashes represent the labeled lines shifted by $\sim10,000$ km s$^{-1}$. We also show a composite HiBAL QSO spectrum from SDSS \citep{SDSS_QSO}. TDE\,2025aarm exhibits both NUV and FUV 10,000 $\rm km~s^{-1}$ blueshifted BALs, similar to iPTF15af or AT\,2019qiz and dissimilar to ASASSN-14li, which shows only transient \textit{emission} features of similar line species. Several of these absorption troughs are accompanied by adjacent emission, producing P~Cygni-like profiles indicative of line formation in an expanding photosphere. This spectrum of TDE\,2025aarm is the earliest UV spectrum of a TDE ever taken, at $\Delta t \approx - 20$ days relative to optical peak light. We mark the rest wavelengths of the detected broad and blueshifted absorption lines in TDE\,2025aarm. $\Delta t$ values for comparison TDEs were taken from \citet{vanVelzen21}. AT\,2019qiz, iPTF\,15af, and ASASSN-14li spectra have been binned for clarity.}
    \label{fig:spec-comp}
\end{figure*}

We obtained a single UV spectrum with the Hubble Space Telescope (HST) Space Telescope Imaging Spectrograph (STIS) (GO-18049; PI E. Hammerstein) on 2025 Nov. 07. It was observed with the $52\arcsec\times 0\farcs2$ aperture. We used the G140L and G230L gratings for the FUV and NUV MAMA detectors, respectively. The FUV and NUV were observed over a single orbit each, with total exposure times of 2569 s in the FUV and 2050 s in the NUV. We use the coadded spectrum resulting from the Hubble Advanced Spectral Products \citep[HASP;][]{HASP} pipeline in our analysis. The spectrum is shown in Figure \ref{fig:spec-comp}. Further epochs were not obtained until late times due to HST visibility constraints and will be presented in future work. All the HST data used in this paper can be found in MAST: \dataset[10.17909/gtj0-4q51]{http://dx.doi.org/10.17909/gtj0-4q51}.

\subsection{Optical/IR Spectroscopy}
\begin{figure*}
    \centering
    \includegraphics[width=0.95\textwidth]{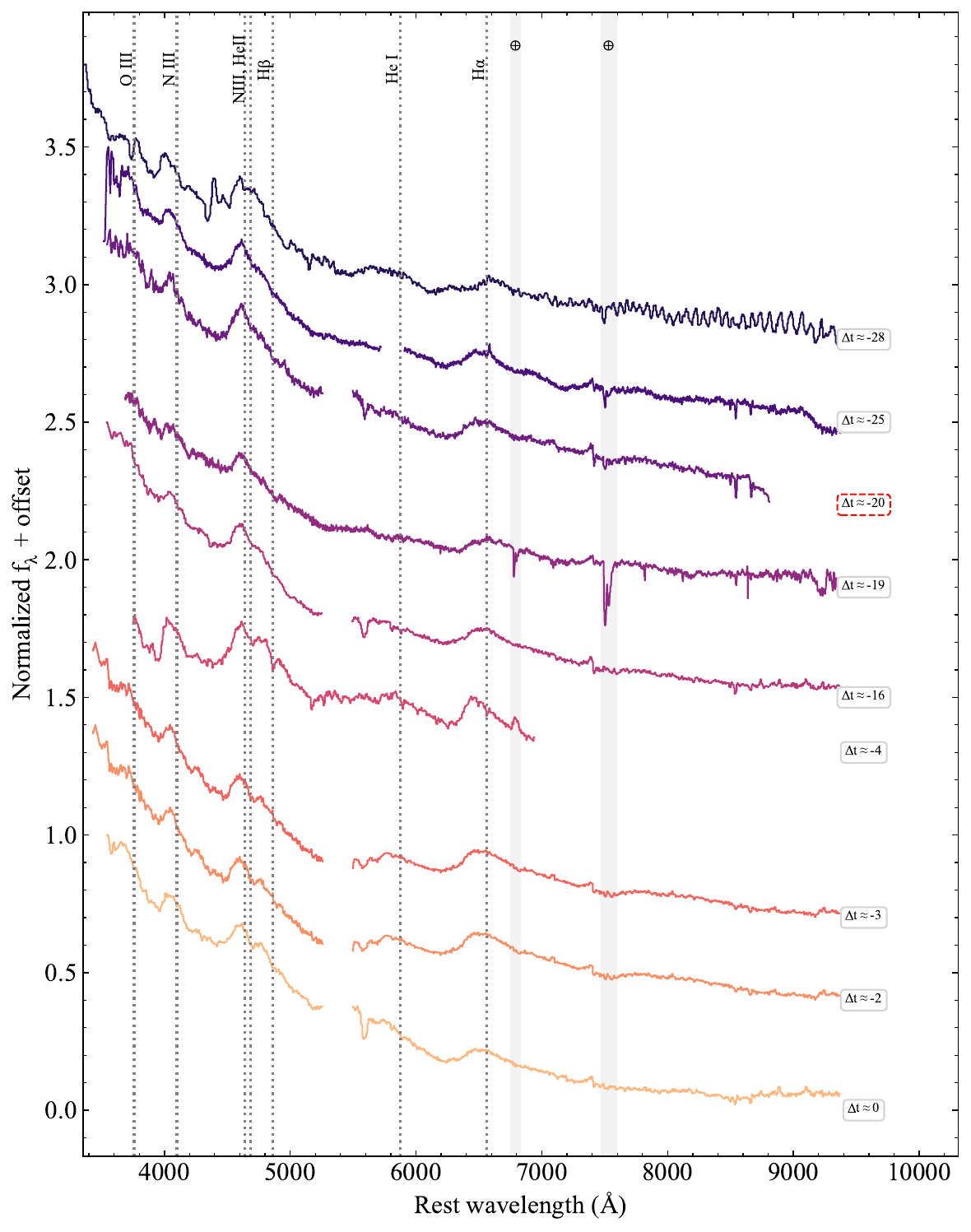}
    \caption{Pre-peak, host-subtracted optical spectra of TDE\,2025aarm spanning from $\Delta t = -28$ days to $\Delta t = 0$ days, in the rest frame relative to optical peak brightness. Broad H$\alpha$, H$\beta$, \ion{He}{2}, and \ion{N}{3} emission is evident from the earliest times. \ion{He}{1} emission is also possible in spectra at early times. The host subtraction outside of the high-resolution 3600--7490 \AA~range in the stellar templates is more uncertain, as can be seen in the \ion{Ca}{2} NIR triplet. We mark the rest wavelengths of commonly observed TDE emission lines. The Kast spectrum taken at the same epoch as the HST/STIS spectrum ($\Delta t \approx -20$ days) is denoted by the dashed red box. We note that the overlap between the red and blue detectors for Kast is at $\lambda_{\rm obs} \approx 5600$ \AA, and we mask pixels in this region. We also mark regions affected by telluric absorption. Spectra are summarized in Table \ref{tab:opt-spec}.}
    \label{fig:opticalspec}
\end{figure*}

\begin{figure*}
    \centering
    \includegraphics[width=0.85\textwidth]{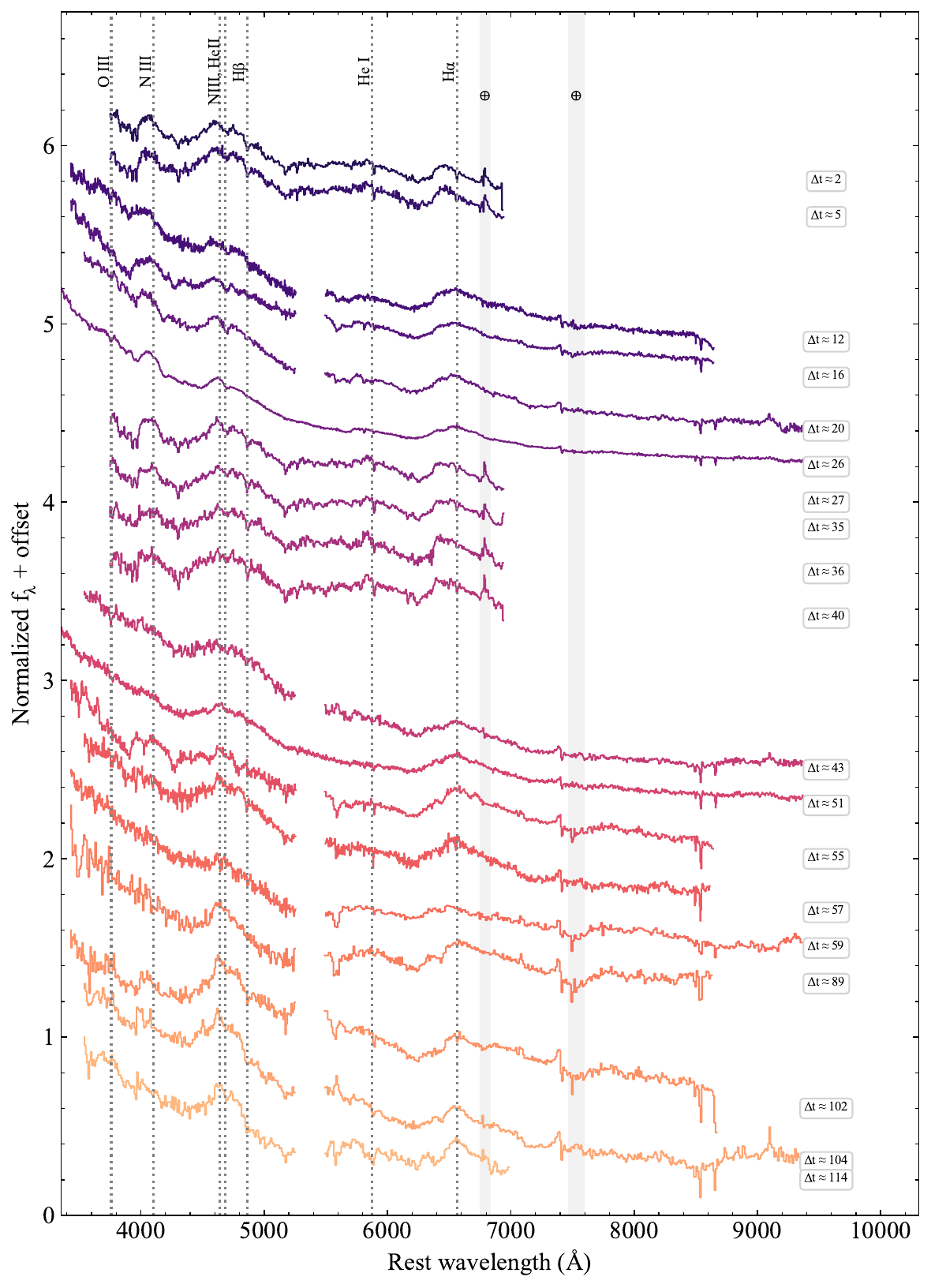}
    \caption{Post-peak, host-subtracted optical spectra of TDE\,2025aarm spanning from $\Delta t = +2$ days to $\Delta t = +114$ days, in the rest frame relative to optical peak brightness. Broad H$\alpha$, H$\beta$, \ion{He}{2}, and \ion{N}{3} emission is persistent throughout. We note that the overlap between the red and blue detectors for Kast is at $\lambda_{\rm obs} \approx 5600$ \AA, and we mask pixels in this region. We also mark regions affected by telluric absorption. Spectra are summarized in Table \ref{tab:opt-spec}.}
    \label{fig:opticalspec2}
\end{figure*}

Here we detail the optical and IR spectra obtained of TDE\,2025aarm listed in Table \ref{tab:opt-spec}. In total, we obtained and present 28 optical/IR spectra from $-25$ days to $+114$ days (rest-frame). We perform host subtraction for all optical spectra using the following procedure, accounting for the differing spectral resolution of each instrument/setup. As a host-galaxy template, we use the resulting spectrum from the \texttt{prospector} fit described in Section \ref{sec:host} and convolve it to the resolution of each instrument/setup as measured from night-sky lines. We then perform a fit to the observed spectrum using a linear combination of the host template normalized by a free constant and a blackbody spectrum, which provides an adequate fit over the range of the spectrum in order to remove the host galaxy contribution. The normalized host galaxy template is subtracted from the observed spectrum and the resulting host-subtracted spectrum is then flux-calibrated to the ZTF $g$-band observation nearest to the date of the observed spectrum, typically to within 2 days. We show all host-subtracted spectra in Figures \ref{fig:opticalspec} and \ref{fig:opticalspec2}.

\begin{deluxetable}{ccr}
\label{tab:opt-spec}
\tablecaption{Spectra of TDE\,2025aarm}
\tablehead{
\colhead{Date} & \colhead{Tel./Inst.} & \colhead{$\Delta t$ (days)}}
\startdata
2025-10-30 & FTN/FLOYDS\tablenotemark{*} & $-28$\\
2025-11-02 & Magellan/IMACS & $-25$\\
2025-11-07 & HST/STIS & $-20$ \\
2025-11-07 & Shane/Kast & $-20$ \\
2025-11-07 & Keck/NIRES & $-20$ \\
2025-11-08 & Magellan/MagE & $-19$\\
2025-11-10 & Shane/Kast & $-16$\\
2025-11-23 & SOAR/GHTS & $-4$ \\
2025-11-24 & Shane/Kast & $-3$\\
2025-11-25 & Shane/Kast & $-2$\\
2025-11-27 & Shane/Kast & $+0$ \\
2025-11-30 & SOAR/GHTS & $+2$ \\
2025-12-03 & SOAR/GHTS & $+5$ \\
2025-12-09 & Shane/Kast & +12\\
2025-12-14 & Shane/Kast & +16\\
2025-12-18 & Shane/Kast & +20 \\
2025-12-23 & Keck/LRIS & +26\\
2025-12-25 & SOAR/GHTS & +27 \\
2026-01-03 & SOAR/GHTS & +35 \\
2026-01-04 & SOAR/GHTS & +36\\
2026-01-08 & SOAR/GHTS & +40 \\
2026-01-11 & Shane/Kast & +43\\
2026-01-19 & Keck/LRIS & +51\\
2026-01-23 & Shane/Kast & +55\\
2026-01-25 & Shane/Kast & +57 \\
2026-01-27 & Shane/Kast & +59\\
2026-02-26 & Shane/Kast & +89\\
2026-03-11 & Shane/Kast & +102\\
2026-03-13 & Shane/Kast & +104\\
2026-03-24 & Shane/Kast & +114\\
\enddata
\tablecomments{The sample of UV through IR spectra included in this paper. We give the rest-frame time relative to peak light in days as $\Delta t$.}
\tablenotetext{*}{From \citet{Faris_Brutusclass}.}
\end{deluxetable}

\subsubsection{Shane/Kast}

We obtained 16 epochs of optical spectroscopy using the Kast spectrograph on the Shane 3\,m telescope at Lick Observatory \citep{millerstone93}. The majority of the Kast observations used a 600/4310 grism on the blue side, D57 dichroic beamsplitter, and the 600/7500 grating on the red side, along with a 2$\arcsec$ slit, resulting in $\sim$5\,\AA\ resolution over the range 3500--8750~\AA. Additional epochs of Kast observations used the 300/7500 grating on the red side, which extended spectral coverage to 10,500~\AA, but at a lower resolution of $\sim$10~\AA. All of the spectra were taken at the parallactic angle to minimize slit losses caused by atmospheric dispersion \citep{Filippenko_PA}. We reduced the spectra with \texttt{pypeit} \citep{pypeit:joss_pub}, following standard reduction procedures for the Kast spectrograph\footnote{\url{https://pypeit.readthedocs.io/en/1.17.4/tutorials/kast_howto.html}}, which includes telluric absorption correction through comparison with a standard star spectrum.

\subsubsection{Keck/LRIS}

In addition, we obtained two spectra with the Low-Resolution Imaging Spectrometer (equipped with an atmospheric dispersion corrector) on the Keck-I 10~m telescope (LRIS; \citealt{lrispaper}). The LRIS observations used the 600/4000 grism on the blue side, the d560 dichroic beamsplitter, and the 400/8500 grating on the red side to cover the full optical range of 3150--10,300\,\AA. With a 1$\arcsec$-wide slit, this resulted in resolutions of $\sim$4 and 6~\AA\ on the blue and red sides, respectively. Data reduction followed standard procedures, including telluric correction, as outlined by \citet{silverman2012}.

\subsubsection{Magellan/IMACS and MagE}
We obtained one epoch of spectroscopy with the Inamori-Magellan Areal Camera and Spectrograph (IMACS) on the Magellan Baade telescope on 2025 Nov. 02 (PI J. Simon). All observations were taken in the F/2 spectroscopic mode, with the FS\_2471 slit mask, the 0\farcs7 slit, and a grating of 400 lines mm$^{-1}$. The spectrum covers 3900--8500 \AA. The IMACS spectrum was reduced via standard IRAF routines,  which include bias subtraction, flatfielding, extraction, and wavelength  calibration with a HeNeAr exposure obtained immediately after the object observation. Flux calibration and telluric feature correction were performed using a spectrum of LTT9491 observed during the same night with the same  instrumental setup.

We obtained one epoch of spectroscopy with the Magellan-Echelette (MagE) Spectrograph on the Magellan Baade telescope on 2025 Nov. 08 (PI B. Mockler). All observations were taken with the 0.85"-wide slit and the standard MagE grating of 175 lines mm$^{-1}$. The MagE spectrum of TDE2025aarm was reduced using standard \texttt{pypeit} routines  \citep{pypeit:joss_arXiv, pypeit:zenodo}, including bias subtraction, flat-fielding, and spectral extraction as well as wavelength calibration with a ThAr exposure obtained immediately after the observation. Flux calibration was performed using the spectra of GD50, a spectrophotometric standard star observed on the same night with the same instrumental setup.

\subsubsection{SOAR/GHTS}
We obtained seven epochs of long-slit spectroscopy of TDE 2025aarm with the Goodman High throughput Spectrograph (GHTS; \citealt{Clemens2004}) mounted on the Southern Astrophysical Research (SOAR) telescope on 2025 Nov. 23 (PI J. Carney), 2025 Nov. 30 (PI J. Carney), 2025 Dec. 03  (PI J. Carney), 2025 Dec. 25 (PI N. Law), 2026 Jan. 03 (PI J. Carney), 2026 Jan. 04 (PI N. Law), and 2026 Jan. 08 (PI N. Law). All observations were taken with a grating of 400 lines mm$^{-1}$ and a $1.0''$-wide slit mask in the M1 spectroscopic setup (hereafter 400M1) with $2 \times 2$ binning using the GHTS Red Camera. The 400M1 spectra cover a wavelength range of 3800--7040 \AA. The spectra were reduced with \texttt{pypeit} \citep{pypeit:joss_arXiv, pypeit:zenodo}, using comparison-lamp spectra taken immediately before and/or after target observation. Flux calibration was performed using observations of HR1996 taken on 2025 Dec. 03 with an identical 400M1 setup and $2 \times 2$ binning. 

\subsubsection{Keck/NIRES}
We also obtained an epoch with the Near-Infrared Echellette Spectrometer \citep[NIRES;][]{nires}, on the Keck-II 10~m telescope (PI M. Kasliwal). Observations were performed on 2025 Nov. 08, and consisted of $4 \times 300$~s exposures using $6''$ dithers in an ABBA pattern. The data were reduced with the standard \texttt{idl} pipeline using \texttt{spextool} \citep{spextool}, with telluric corrections performed using \texttt{xtellcor} \citep{xtellcor}.  


\subsection{Chandra/ACIS-S (0.3--10 keV)}

Chandra observed TDE\,2025aarm three times with the Advanced CCD Imaging Spectrometer-S (ACIS-S), beginning on 2025 Nov. 05 at 22:53:52 (24.75 ks, ObsID 31982, PI J. Somalwar), 2025 Dec. 22 at 16:35:22 (15.03 ks, ObsID  32075, PI J. Somalwar), and 2026 Apr. 11 at 18:25:04 (14.74 ks,  ObsID 32275, PI  Wichern). We analyzed the data using \texttt{CIAO} v4.17 package as recommended, including reprocessing with the \texttt{chandra\_repro} task. Background light curves were extracted to ensure that there are no times affected by solar flares or any other significant background. We refined the position of AT\,2025aarm in each observation by centroiding, extract a spectrum  within the 95\% encircled counts fraction radius, and apply a point-source aperture correction. An annulus from 5--30 times the source-extraction radius was used to measure the background. 
We find total, background-subtracted count rates of $6.96^{+3.51}_{-2.57}\times10^{-4}\ \mathrm{c\,s^{-1}}$ (ObsID 31982), $3.63^{+0.92}_{-0.77}\times10^{-3}\ \mathrm{c\,s^{-1}}$ (ObsID 32075), and $7.36^{+1.28}_{-1.12}\times10^{-3}\ \mathrm{c\,s^{-1}}$ (ObsID 32275). 

The extracted spectra were fit with the \texttt{sherpa} code to an absorbed power-law model (\texttt{phabs*zpowerlw}) with the absorption fixed to the Milky Way value $N_{\rm H} = 3.98\times10^{20}$\,cm$^{-2}$ \citep{HI4PI16}. This model was fit to each epoch independently with a Markov Chain Monte Carlo (MCMC), running the chains until the Gelman-Rubin statistic reached the threshold $<1.01$. We extracted the unabsorbed flux and spectral indices from the resulting posterior samples. The best-fit unabsorbed fluxes in the 0.3--10\,keV band are $1.99^{+0.61}_{-0.48}\times10^{-14}\ \mathrm{erg\,cm^{-2}\,s^{-1}}$ (ObsID 31982), $9.35^{+1.52}_{-1.32}\times10^{-14}\ \mathrm{erg\,cm^{-2}\,s^{-1}}$ (ObsID 32075), and $1.87^{+0.19}_{-0.18}\times10^{-13}\ \mathrm{erg\,cm^{-2}\,s^{-1}}$ (ObsID 32275). The best-fit spectral indices are $1.96^{+0.53}_{-0.51}$ (ObsID 31982), $1.97^{+0.29}_{-0.31}$ (ObsID 32075), $1.63^{+0.21}_{-0.19}$ (ObsID 32275). These spectral indices are consistent within uncertainties, so we improve the spectral slope constraint by simultaneously fitting the three spectra with fixed spectral slope and independent amplitudes. The best-fit joint spectral slope is $\Gamma = 1.75^{+0.15}_{-0.14}$.

\subsection{NuSTAR  (3--79 keV)} \label{SubSec:NuSTAR}
The Nuclear Spectroscopic Telescope (NuSTAR) started observing AT\,2025aarm  on 2025 Nov. 11 at 11:16:09 ($\delta t\approx 16$\,d before optical light curve peak), ObsID 91191646002, PI R. Margutti). We analyzed the data with the NuSTAR Data Analysis Software (v2.1.2) and calibration files (version 20240104). A first extraction with \texttt{nupipeline} and standard filtering provided no statistical evidence for emission at the source location. However, our analysis shows that the observations are affected by a large episode of flaring of the background due to solar activity.  We removed the interval of times affected by solar flares and significant radiation-belt backgrounds using standard background plots and custom \texttt{python} scripts by redefining the Good Time Intervals (GTIs) of extraction of our products. Our final cleaned event files have exposure times of 34.4\,ks and 34.7\,ks for modules A and B, respectively. 

We do not find evidence for statistically significant emission at the location of the TDE. Combining signal from both modules, we report a limit of $4.3\times 10^{-4}\,\rm{counts\,s^{-1}}$ ($3\sigma$ c.l.) in the 8--20 keV energy band (which is around the peak of the NuSTAR effective area). For an assumed nonthermal $F_{\nu}\propto \nu^{-1}$ spectrum that is typical of Comptonization components, the  count-rate limit translates into a flux limit of $2.4\times 10^{-14}\,\rm{erg\,s^{-1}cm^{-2}}$ (8--20 keV). 

\section{Host Galaxy} \label{sec:host}
\begin{figure}
    \centering
    \includegraphics[width=\linewidth]{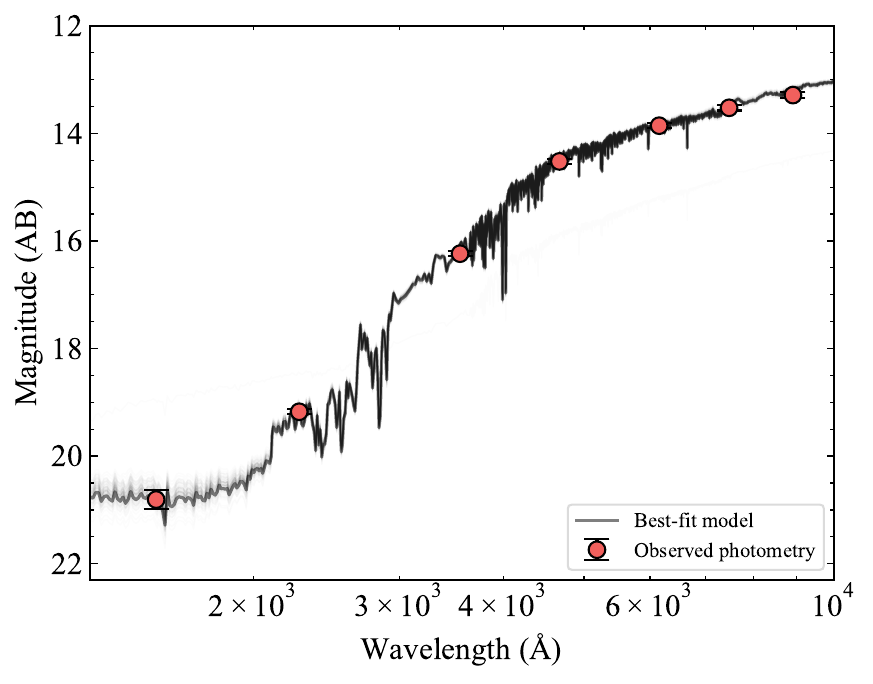}
    \caption{Results from our \texttt{prospector} modeling of the host galaxy of TDE\,2025aarm. We show the best-fit model as the darker gray line and the observed GALEX FUV and NUV photometry, as well as the observed SDSS $ugriz$ photometry denoted by the circles.}
    \label{fig:host-fit}
\end{figure}

We follow the methodology of \citet{vanVelzen21} and \citet{Hammerstein23} to fit the pre-flare host spectral energy distribution (SED) of TDE\,2025aarm in order to obtain host-galaxy properties, host photometry to use for Swift/UVOT subtraction, and a host spectrum for constructing host-subtracted transient spectra. The SED includes fitting of Sloan Digital Sky Survey (SDSS) $ugriz$ model magnitudes and GALEX NUV and FUV photometry \citep{martin05}. We model the SED using the software package \texttt{prospector} \citep{Johnson2021}, which uses the \texttt{Flexible Stellar Population Synthesis} (FSPS) code \citep{Conroy2009} to construct a spectrum from simple stellar populations. The model consists of five free parameters, including the stellar mass, the \citet{Calzetti00} dust model optical depth, stellar population age, metallicity, and $e$-folding time of the star-formation history ($\tau_{\rm sfh}$). We adopt the same parameter choices as \citet{Mendel14} to match the procedure carried out by \citet{vanVelzen21} and \citet{Hammerstein23}. We fit the SED using \texttt{emcee}, employing 100 walkers, running for 20,000 steps ($\gg50$ times the autocorrelation time to ensure convergence). We discard the first 2,000 steps as burn-in. The resulting host spectrum is compared to the observed host photometry in Figure \ref{fig:host-fit}, and the resulting host-galaxy properties are given in Table \ref{tab:host}. We used the resulting model optical spectrum to perform host subtraction for the transient spectra.

The black hole mass implied by the host-galaxy stellar mass from \texttt{prospector} is $\log(M_{\rm BH}/M_\odot) = 6.69$, using the stellar mass vs. SMBH mass relation of \citet{reines15}. This is within the uncertainties of the black hole mass inferred from the velocity dispersion measured from the DESI spectrum fit using \texttt{FastSpecFit}, which is $\log(M_{\rm BH}/M_\odot) = 6.98 \pm 0.46$ for $\sigma_\star = 107.84 \pm 2.23$ km s$^{-1}$, if using the $M_{\rm BH}-\sigma_\star$ relation from \citet{gultekin09}.

\begin{deluxetable}{cc}
\label{tab:host}
\tablecaption{Host-galaxy properties from \texttt{prospector}}
\tablehead{
\colhead{Parameter} & \colhead{Value}}
\startdata
$\log(M/M_\odot)$ & $10.13_{-0.08}^{+0.07}$ \\
$E(B-V)$ (mag) & $0.05_{-0.02}^{+0.01}$ \\
$\log(\rm age/Gyr)$ & $6.02_{-1.48}^{+1.55}$ \\
$\log(\rm \tau_{\rm sfh}/Gyr)$ & $0.70_{-0.17}^{+0.18}$ \\
$\log(Z/Z_\odot)$ & $-0.09_{-0.10}^{+0.08}$ \\
\enddata
\tablecomments{Host-galaxy properties derived from \texttt{prospector} fits to the pre-flare host photometry, including the total stellar mass, color excess, age of the galaxy, the star formation e-folding timescale, and the metallicity.}
\end{deluxetable}

\section{Analysis and Results} \label{sec:results}

\subsection{Light-curve Evolution} \label{SubSec:lcfit}
For comparison to previous TDEs, we characterize the light-curve evolution by the time it takes the TDE to rise from half-maximum brightness to maximum, and then decay from maximum to half-maximum ($t_{\rm 1/2, rise}$ and $t_{\rm 1/2, decay}$, respectively). We first fit for the time of peak $g$-band flux ($t_{\rm peak}$) and the flux at peak ($f_{\rm peak}$) using a second-order polynomial in a window of 50 days surrounding the maximum $g$-band flux measurement. We then estimate $t_{\rm 1/2, rise}$ and $t_{\rm 1/2, decay}$ by linearly interpolating the $g$-band light curve. To estimate the uncertainties on these parameters, we perform a Monte Carlo with 600 realizations of the light curve, sufficient such that the Monte Carlo sampling error is negligible relative to the measurement uncertainties. We find that $t_{\rm peak} = {\rm MJD}~61006.54_{-1.01}^{+1.00}$ days, which we use as the reference time for the remaining analysis, $t_{\rm 1/2, rise} = 31_{-1}^{+1}$ days, and $t_{\rm 1/2, decay} = 51_{-3}^{+3}$ days, in the rest frame.

The $t_{\rm 1/2, rise}$ and $t_{\rm 1/2, decay}$ timescales are well matched to previous samples of TDEs. The sample of TDEs presented by \citet{Yao2023} had mean $t_{\rm 1/2, rise}$ and $t_{\rm 1/2, decay}$ timescales of $\sim20$ days and $\sim37$ days, respectively. While the $t_{\rm 1/2, rise}$ and $t_{\rm 1/2, decay}$ for TDE\,2025aarm are slightly longer at $\sim31$ and $\sim 51$ days, they are well within ranges found for previous TDEs.

\subsection{SED Fitting}
We fit SEDs of TDE\,2025aarm at multiple epochs from pre- to post-peak to estimate the properties and evolution of the UV--optical blackbody over the course of the light curve. Each SED spans a time bin of less than two days, selected to include at least two UV photometric measurements from UVOT together with the optical photometry from ZTF and ATLAS. This results in 21 epochs beginning at $\Delta t \approx -26$ days until $\Delta t \approx +72$ days, rest frame.

\begin{figure}
    \centering
    \includegraphics[width=0.9\linewidth]{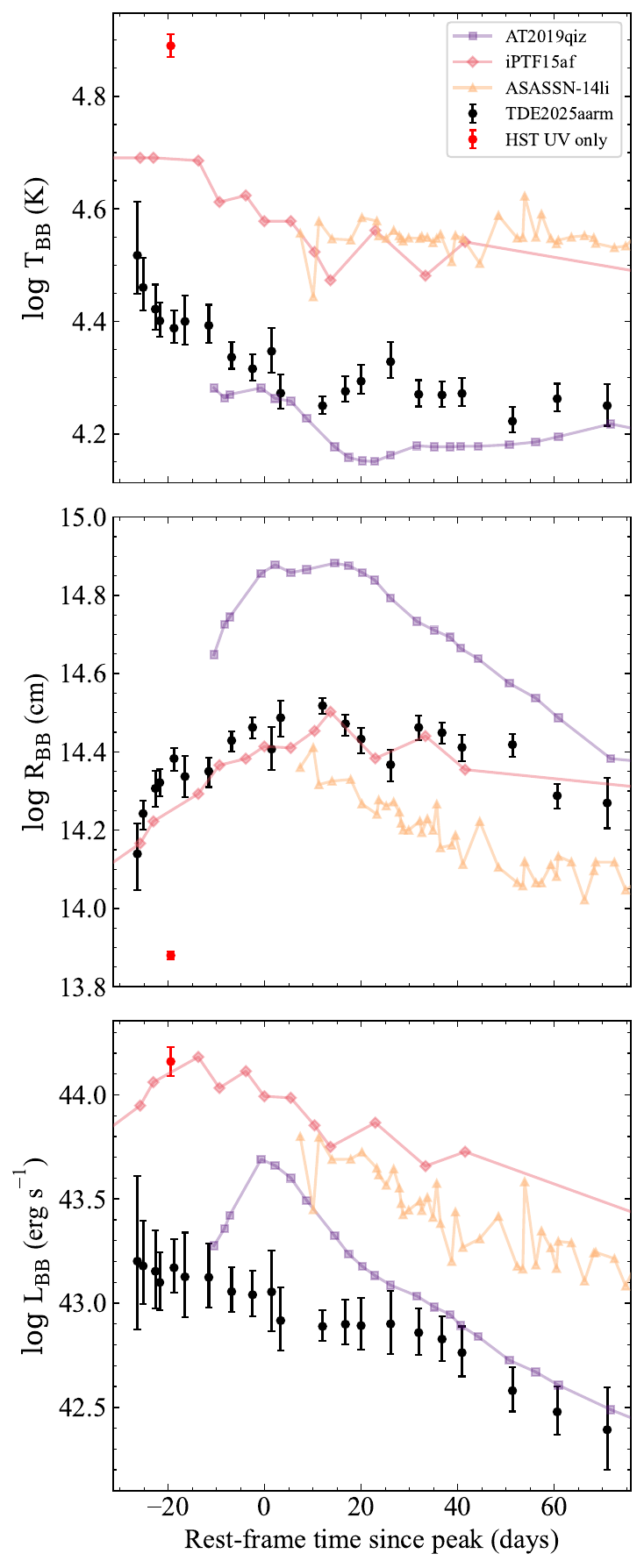}
    \caption{Temperature, radius, and luminosity evolution from fitting a blackbody to SEDs of TDE\,2025aarm. We include the results from fitting a blackbody to the HST/STIS continuum as the red point. Similar fits are included to ASASSN-14li, iPTF\,15af, and AT\,2019qiz. TDE\,2025aarm shows similar evolution to previous TDEs, with a higher temperature pre-peak that stabilizes to a lower temperature post-peak, resulting in a bolometric luminosity that peaks before the optical $g$ band.}
    \label{fig:SED_evol}
\end{figure}

We model the continuum as a single blackbody in order to make direct comparisons to previous works, but we caution that this may not be truly representative of the full SED (see Section \ref{SubSec:continuum-fit}). To estimate the blackbody temperature and radius and thereby the blackbody luminosity, we employ an MCMC using \texttt{emcee} \citep{Foreman-Mackey2013} with 100 walkers and a minimum number of steps greater than or equal to 50 times the longest autocorrelation time ($\tau_{\rm corr}$) for the estimated parameters to ensure convergence. We discard $\sim3\times\tau_{\rm corr}$ as burn-in. Flat priors are used for all parameters with $10^4 < T_{\rm BB}/\rm K < 10^6$ and $10^{13} < R_{\rm BB}/\rm cm < 10^{16}$, and  a ``white-noise'' term is included to account for additional variance in the data not captured by the formal measurement uncertainty. The results of these SED fits are illustrated in Figure \ref{fig:SED_evol}, which shows the evolution in $T_{\rm BB}$, $R_{\rm BB}$, and $L_{\rm BB}$. For comparison, we also plot the same TDEs as in Figure \ref{fig:spec-comp} with similarly early, well-sampled UV and optical light curves, including ASASSN-14li \citep[SED fits from][]{Hung2017}, iPTF\,15af \citep{Blagorodnova2019}, and AT\,2019qiz \citep{Hung2021}.

The evolution of TDE\,2025aarm is similar to that of previous TDEs. The blackbody temperature is higher pre-optical peak before cooling slightly and stabilizing at a lower temperature post-peak, which is also seen in iPTF\,15af and AT\,2019qiz. This is consistent with the findings of \citet{Aamer2026}, though \citet{Aamer2026} note subtle deviations in the blackbody parameters that correspond to deviation from smooth evolution in the light curve and $>5, 000-10,000$ K jumps in the blackbody temperature. The bolometric luminosity peaks earlier than the optical $g$-band peak by at least 20 days. This is similarly seen in iPTF\,15af, where the bolometric luminosity peaks $\sim 18$ days before optical peak. We also show the fit to the NUV continuum in the UV spectrum (see Section \ref{SubSec:continuum-fit}), which results in a much higher blackbody temperature and luminosity and smaller blackbody radius than is estimated from the UV--optical photometry. The blackbody luminosity inferred from the optical photometry is consistent with previous samples of TDEs; for example, \citet{Hammerstein23} find that most TDEs peak in the range $10^{43-44}$ erg s$^{-1}$. While the blackbody luminosity inferred from the UV continuum is higher, it is still consistent with previous samples of TDEs, though we emphasize that Figure \ref{fig:continuum-fit} demonstrates that the continuum deviates significantly from a single blackbody. The fits to the UV and optical continuum from the spectra are discussed in the next section.

\subsection{UV--Optical Continuum Fitting} \label{SubSec:continuum-fit}
Previous studies of TDEs largely rely on photometry to estimate the properties of the UV--optical continuum \citep[e.g.,][]{vanVelzen21, Hammerstein23}, primarily by fitting a single-temperature blackbody. When UV spectra are available \citep[e.g.,][]{Blagorodnova2019, Hung2021, Yao_lhc_kmq, Zhu26}, the discrepancy between the estimated single-temperature blackbody continuum and the true continuum, particularly in the FUV, becomes clear.

Motivated by this, we fit the UV--optical continuum of TDE\,2025aarm with several models, including a blackbody and a power law, to quantify this discrepancy, particularly in the resulting bolometric luminosity. For each of the fits, we employ the same method and priors as for the blackbody photometry fits using \texttt{emcee}. For the power-law fit, we use $2 < \alpha < 6$ for the power-law spectral index, where $f_\lambda \propto \lambda^{-\alpha}$. For all fits, we mask strong host absorption, as well as TDE emission and absorption lines.

\begin{figure*}
    \centering
    \includegraphics[width=\linewidth]{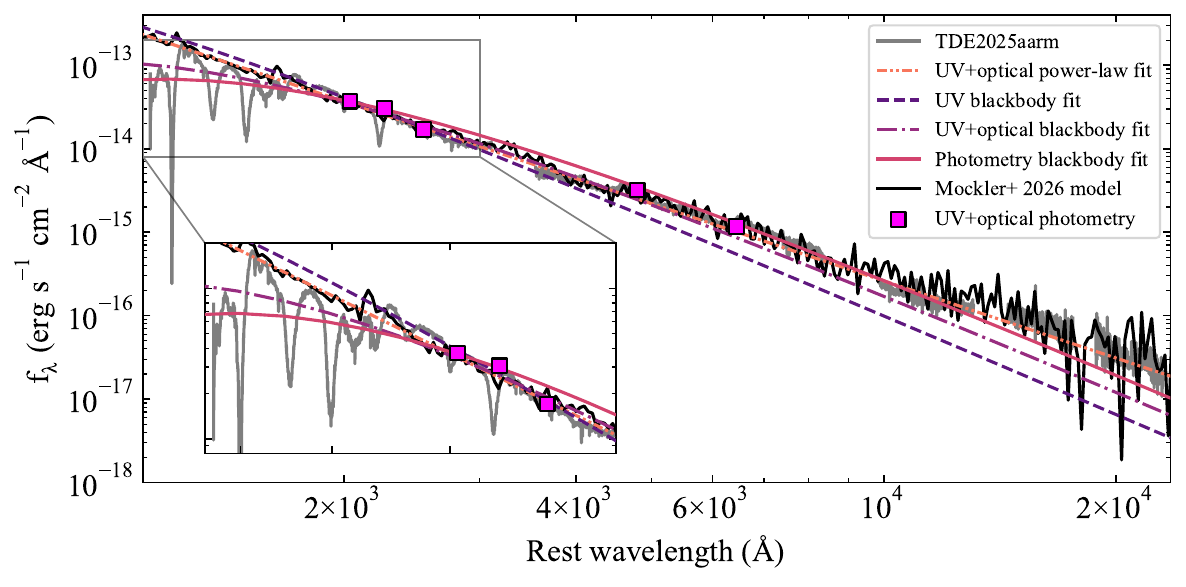}
    \caption{Fits to the UV--optical continuum using the UV and optical spectra or the UV and optical photometry. We include the IR spectrum for comparison, but do not use it in the fits. The UV+optical and photometry fits are largely consistent with one another, but the UV blackbody fit places the peak of the blackbody at much shorter wavelengths. In all cases except the power-law fit, the optical--IR continuum is underestimated.}
    \label{fig:continuum-fit}
\end{figure*}

The fits, shown in Figure \ref{fig:continuum-fit} with corresponding parameter values in Table \ref{tab:continuum-fit}, are as follows. We first fit the UV and optical spectra from HST and Kast with a single blackbody. In the FUV, the regions of usable continuum are sparse and difficult to select, so we instead only fit line-free regions in the NUV, above $\sim 1700$ \AA. This fit results in blackbody parameters comparable to those obtained from fits to the photometry, with $\log(T_{\rm BB}/\rm K)=4.512 \pm 0.004$ and $\log(R_{\rm BB}/\rm cm)=14.22 \pm 0.004$, and bolometric luminosities consistent within uncertainties. Notably, this fit significantly underestimates regions of absorption in the UV and the longer wavelengths in the optical. We then repeat this fit but only using the NUV continuum. This results in a much better estimate of both the NUV and FUV continuum, but an underestimate of the optical continuum. In all cases, the IR continuum (which is not used in the fit) is underestimated. Lastly, we fit a power law to the UV and optical continuum. This results in the best match to the continuum from UV to IR, despite not including the IR spectrum in the fit, but it is much shallower than the expected $f_\lambda \propto \lambda^{-4}$ Rayleigh-Jeans tail, with $\alpha=3.08\pm0.01$.

\begin{figure*}
    \centering
    \includegraphics[width=\linewidth]{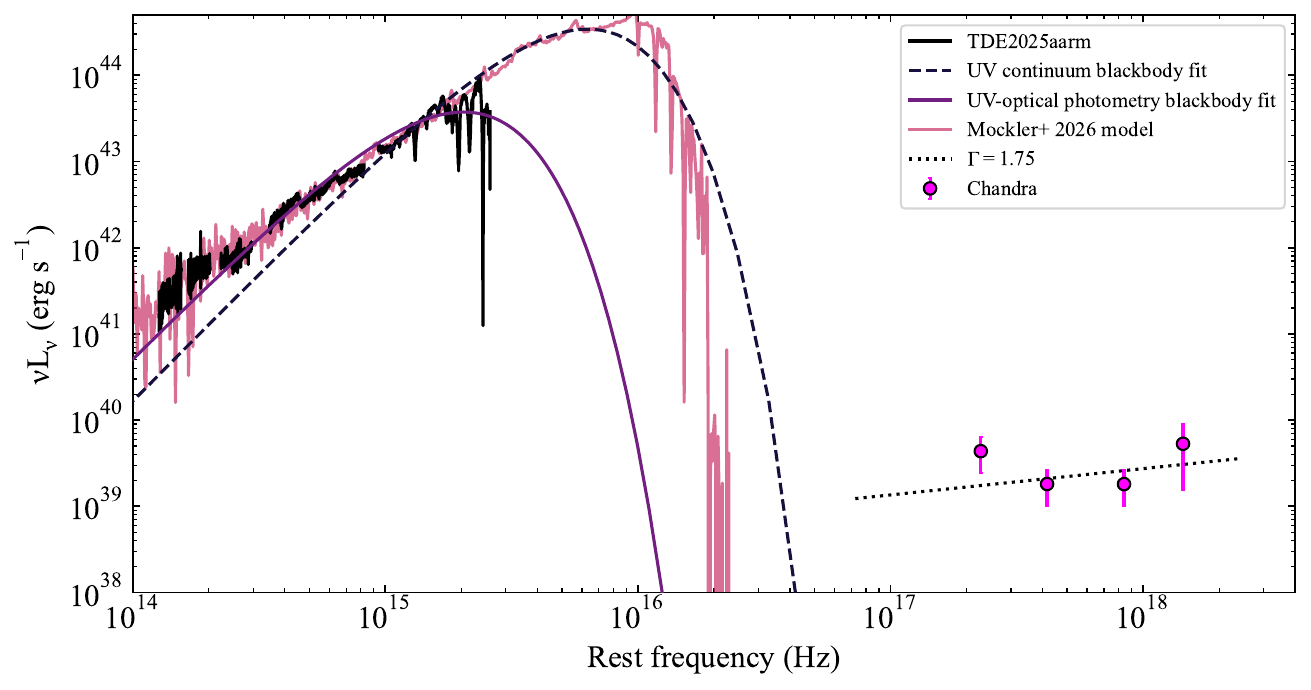}
    \caption{SED of TDE\,2025aarm from the IR to X-rays at $\Delta t \approx -20$~d before optical light curve peak, showing the blackbody fits to the UV continuum and to the UV and optical photometry. The difference between the two fits to the UV and optical spectra and photometry is clear, the fit to the photometry representing a bolometric luminosity that is $\lesssim 85\%$ of the bolometric luminosity estimated from the UV continuum alone. In both cases, the IR continuum is underestimated, suggesting an excess in the NIR and optical.}
    \label{fig:superSED}
\end{figure*}

The resulting bolometric luminosities from the various blackbody fits are starkly different. The bolometric luminosity estimated only from the UV continuum is a factor of $\sim$9 higher than that estimated from the blackbody fits to either the UV+optical spectrum or photometry alone. It is clear that single-blackbody approximations for the UV and optical continuum miss a significant fraction of energy emitted at shorter wavelengths. This is illustrated on a larger scale in Figure \ref{fig:superSED}, which includes shorter wavelengths out to X-rays. Previous works \citep[e.g.,][]{Holoien16_14li, Short2020} have shown that \ion{He}{2} luminosities observed in other TDEs, which require the presence of EUV photons, may be inconsistent with the continuum predicted by the UV/optical photometry. We discuss the source of this discrepancy, particularly as it arises from an electron-scattering dominated atmosphere, and implications in Section \ref{SubSec:SED-discussion}.

\begin{deluxetable*}{cccccc}
\label{tab:continuum-fit}
\tablecaption{Results from UV--optical continuum fitting}
\tablehead{
\colhead{Model} & \colhead{$\log(T_{\rm BB}/\rm K)$} & \colhead{$\log(R_{\rm BB}/\rm cm)$} & \colhead{$\log(L_{\rm BB}/\rm erg~s^{-1})$} & \colhead{$\alpha$} & \colhead{$L_{\rm BB}/L_{\rm BB, UV}$}}
\startdata
  UV blackbody & $4.89^{+0.02}_{-0.02}$ & $13.88^{+0.01}_{-0.01}$ & $44.16^{+0.07}_{-0.07}$ & \nodata & 1.0 \\
  UV+optical blackbody & $4.51^{+0.00}_{-0.00}$ & $14.22^{+0.00}_{-0.00}$ & $43.31^{+0.02}_{-0.02}$ & \nodata & 0.14 \\
  Photometry blackbody & $4.40^{+0.05}_{-0.05}$ & $14.38^{+0.06}_{-0.06}$ & $43.20^{+0.24}_{-0.22}$ & \nodata & 0.11 \\
  UV+optical power-law & \nodata & \nodata & \nodata & $3.08^{+0.01}_{-0.01}$ & \nodata
\enddata
\tablecomments{The results from the UV--optical continuum fitting, including those to the spectra and photometry. We also include the single fit to the UV--optical spectra using a power-law continuum. The final column shows the ratio between the bolometric luminosity of that fit and the bolometric luminosity from the UV-estimated blackbody.}
\end{deluxetable*}

\subsection{UV Broad Absorption Lines}
The UV spectrum displays clear, strong BALs in both the NUV and FUV blueshifted by $\sim$10,000 km\,s$^{-1}$. We find good similarity to previously observed BAL TDEs shown in Figure \ref{fig:spec-comp}, including iPTF\,15af \citep{Blagorodnova2019} and AT\,2019qiz \citep{Hung2021}, which are shown. These are in contrast to objects like ASASSN-14li \citep{cenko16}, which shows only broad emission lines centered around the rest wavelengths of the blueshifted FUV absorption lines in other TDEs. To identify the ions associated with the lines in TDE\,2025aarm, we compare the spectrum to these other TDEs.

In the NUV, we find that there is evidence for weak \ion{Mg}{2} at approximately the same velocity as the stronger BALs. There is possible evidence for another weak BAL at $\sim2460$ \AA, which may be associated with ions such as \ion{Si}{2}, \ion{Ti}{2}, and \ion{C}{2} as identified by \citet{Yan2017} in the SLSN Gaia\,16apd, but this is highly dependent on the continuum in this region. We identify the line at $\sim2226$ \AA~as \ion{C}{3} $\lambda$2297. \citet{Hung2021} identified a similar feature in AT\,2019qiz as \ion{Fe}{2} (UV2) instead of \ion{C}{3}, but this identification would place that line in TDE\,2025aarm at a much higher velocity ($\gtrsim 18,000$ km\,s$^{-1}$) than the other lines seen in either the NUV or FUV. We therefore prefer the identification of this line as \ion{C}{3}. We identify the line at $\sim2006$ \AA~as \ion{Fe}{3} and the line at $\sim 1802$ \AA~as \ion{Al}{3}. In the FUV, we identify very weak Ly$\alpha$ $\lambda1216$ at $\sim1171$ \AA, likely \ion{N}{5} $\lambda12406$ at $\sim1199$ \AA, \ion{O}{1} $\lambda1302$ at $\sim1259$ \AA, strong \ion{Si}{4} $\lambda1402$ at $\sim1354$ \AA, and strong \ion{C}{4} $\lambda1548$ at $\sim1495$ \AA. Note that the blueshifted \ion{N}{5} BAL is in a region contaminated by foreground Ly$\alpha$ absorption. These lines are consistent with what has been found in previous TDEs.

The spectrum also reveals two new BALs that have not yet been noted or identified in TDEs prior to TDE\,2025aarm, with absorption minima at $\sim$1587 \AA~and $\sim$1670 \AA. The bluer line may have contributions from \ion{He}{2}, as this identification would place the blueshifted BAL at $\sim9900$ km\,s$^{-1}$, similar to the other NUV and FUV lines. 
The redder line would largely be inconsistent with blueshifted \ion{N}{3}] absorption, as that would place the BAL at the highest velocity ($\sim 13,700$ km\,s$^{-1}$) of any of the identified lines and more fundamentally, as a semi-forbidden transition it would not be expected to appear in absorption. Interestingly, these lines may appear very weakly in iPTF\,15af (see Figure \ref{fig:spec-comp}), though the determination of the continuum in that source likely prevented firm confirmation and instead \ion{He}{2} and \ion{N}{3}] emission were identified by \citet{Blagorodnova2019}.

Alternatively, \citet{Aspegren2026prep} run one-dimensional radiative-transfer calculations in local thermodynamic equilibrium (LTE) and conduct a survey of different density profiles, velocity structures, and compositions for the stellar debris to model the UV emission and compare to observed TDE spectra. In their model spectra that show approximately similar features to TDE\,2025aarm, the $\sim1587$ \AA~and $\sim1670$ \AA~lines may also have contributions from \ion{Fe}{4} blends and \ion{N}{4} $\lambda 1718$, respectively.

To derive the properties of the broad UV absorption lines, we fit each line with a linear combination of the local continuum and a Gaussian absorption profile. We mask narrow host and foreground absorption lines within the region of the fit. The results of these fits, including the shift and full width at half-maximum intensity (FWHM) of each line, are listed in Table \ref{tab:broad-uv}. The average centroid velocity of all lines is $\sim9600$ km\,s$^{-1}$, with the highest centroid velocity in the \ion{Fe}{3} line ($\sim$10,500 km\,s$^{-1}$) and the lowest centroid velocity in the \ion{Al}{3} line ($\sim8820$ km\,s$^{-1}$).

\begin{figure}
    \centering
    \includegraphics[width=\linewidth]{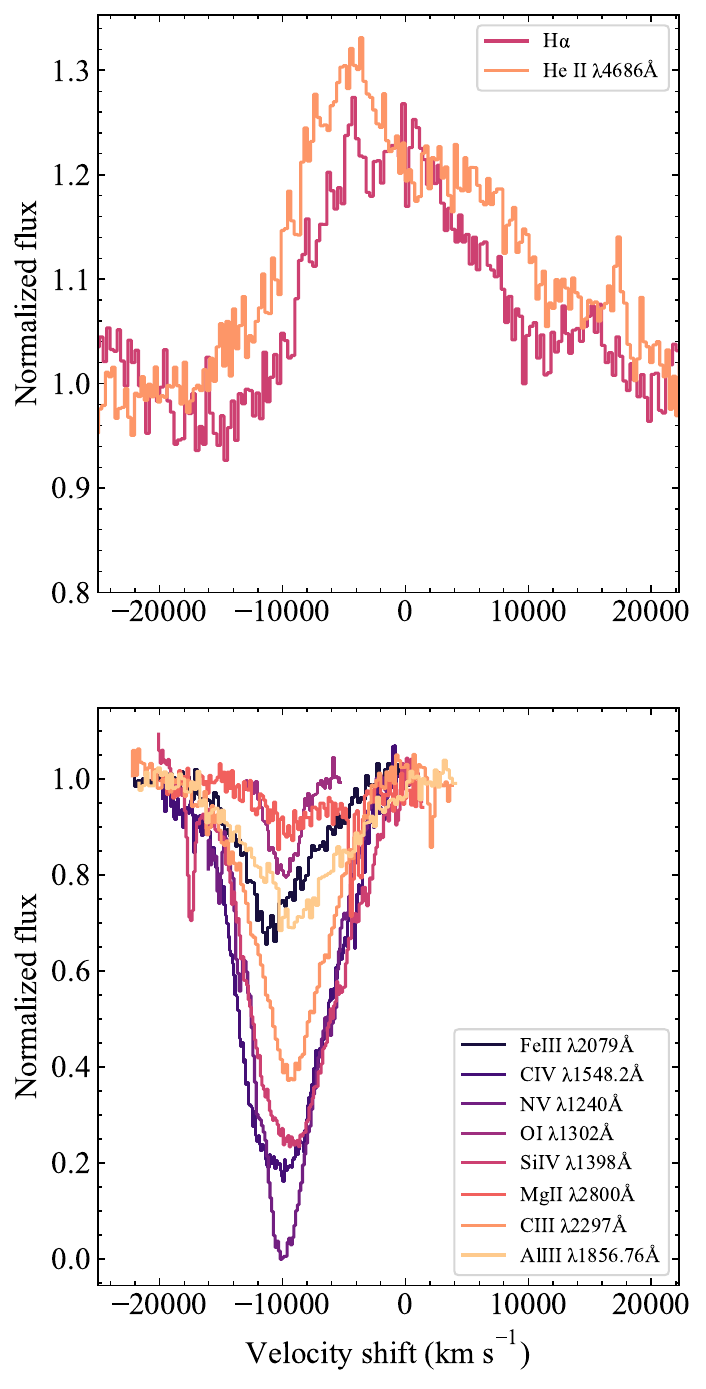}
    \caption{\textit{Top:} H$\alpha$ and \ion{He}{2} profiles from the $\Delta t \approx -20$ day optical spectrum, which was taken within a day of the HST spectrum. There may be weak evidence for P~Cygni-like profiles in the optical lines. The flux is normalized using the local continuum. \textit{Bottom:} Comparison of velocity shifts of detected BALs in TDE\,2025aarm, with flux normalized using a local linear continuum. The NUV and FUV absorption lines are systematically blueshifted by $\sim$10,000 km s$^{-1}$, providing strong evidence for a fast outflow. The legend is ordered from largest to smallest velocity shift. There is a spread in velocity shifts of the lines, with many of the NUV lines exhibiting lower velocity shifts. We exclude Ly$\alpha$ as the continuum in the line region is highly uncertain. }
    \label{fig:line_overlap}
\end{figure}

We compare the line profiles for the UV BALs in Figure \ref{fig:line_overlap}. There is a tentative weak anticorrelation between the line centroid velocity and the line rest wavelength, with many of the FUV lines at higher velocities than the NUV lines. The \ion{Fe}{3} line is a clear outlier among the NUV lines. The shapes of the FUV and NUV lines differ significantly. This is illustrated in Figure \ref{fig:line_overlap}, where the \ion{C}{4} line presents a much broader profile and at a higher velocity than the narrower \ion{C}{3} line. It is likely that a blend of several lines affects the shape of the \ion{C}{4} and other FUV lines. We also show the H$\alpha$ and \ion{He}{2} optical emission lines in the bottom panel of Figure \ref{fig:line_overlap}. The H$\alpha$ profile is broad and slightly blueshifted with FWHM $\approx$ 13,000 km s$^{-1}$. This is consistent with the results of \citet{Simongini_Brutus} and \citet{Baldini_Brutus}. Because the \ion{He}{2} profile is blended with both H$\beta$ and \ion{N}{3}, the precise width and blueshift is difficult to measure, but appears consistent with H$\alpha$ in Figure \ref{fig:line_overlap}.

\begin{deluxetable}{crr}
\label{tab:broad-uv}
\tablecaption{Broad UV absorption lines detected in TDE\,2025aarm}
\tablehead{
\colhead{Ion} & \thead{Velocity \\ ($10^3$ km s$^{-1}$)} & \thead{FWHM \\ ($10^3$ km s$^{-1}$)}}
\startdata
\ion{O}{1} $\lambda 1302$ & $-9.83_{-0.04}^{+0.04}$ & $3.06_{-0.13}^{+0.14}$ \\ 
\ion{Mg}{2} $\lambda2800$ & $-9.25_{-0.17}^{+0.11}$ & $5.09_{-0.85}^{+0.83}$ \\ 
\ion{Al}{3} $\lambda1857$ & $-8.82_{-0.09}^{+0.08}$ & $10.2_{-0.46}^{+0.47}$ \\ 
\ion{Fe}{3} $\lambda 2079$ & $-10.51_{-0.06}^{+0.06}$ & $6.66_{-0.23}^{+0.23}$ \\
\ion{Si}{4} $\lambda 1398$ & $-9.41_{-0.03}^{+0.03}$ & $8.32_{-0.12}^{+0.15}$ \\ 
\ion{C}{3} $\lambda2297$ & $-9.25_{-0.04}^{+0.03}$ & $7.19_{-0.17}^{+0.15}$ \\ 
\ion{C}{4} $\lambda1548$ & $-10.19_{-0.03}^{+0.03}$ & $8.62_{-0.11}^{+0.11}$ \\
\ion{N}{5} $\lambda 1240$ & $-9.86_{-0.04}^{+0.04}$ & $5.34_{-0.39}^{+0.38}$

\enddata
\tablecomments{
Lines are ordered from lower to higher ionization potentials. We note that many lines are likely blends and the line velocity is reported with respect to the wavelength listed in the first column.}
\end{deluxetable}

\section{Discussion} \label{sec:discussion}

\subsection{Blackbody Evolution}
We now place our analysis of TDE\,2025aarm in the context of previous TDEs, particularly those where BALs have been observed, and discuss the implications of their evolution. Figure \ref{fig:SED_evol} shows the evolution of the blackbody parameters for TDE\,2025aarm, AT\,2019qiz, iPTF\,15af, and ASASSN-14li. Both the reprocessing and stream-collision models make predictions for the temperature evolution of UV/optical blackbody pre- to post-optical peak. In the case where circularization of the tidal debris is prompt and outflows or winds driven by super-Eddington accretion are responsible for producing the emission, \citet{Bu2022} found that the blackbody temperature of the outflow should cool rapidly in the first ten days after the peak of the optical light curve, driven by the expansion of the photosphere. This is consistent with what is seen in AT\,2019qiz, but the cooling in both TDE\,2025aarm and iPTF\,15af occurs pre-peak. In contrast, \citet{Steinberg2022} found that in the scenario where the debris-shocking process produces the UV/optical emission, the temperature is expected to cool over the twenty days leading up to optical light-curve peak because the shock-heated debris expands and its photosphere grows and cools during the rise. The temperature then stabilizes around peak once circularization becomes efficient and accretion begins. This cooling pre-peak is consistent with what is seen in iPTF\,15af and TDE\,2025aarm. However, this model also predicts that outflows will appear post-peak when debris circularization is efficient and accretion of the tidal debris begins. In the case of TDE\,2025aarm, strong, fast outflows are already observed pre-peak.

The delay between the peak bolometric luminosity and the peak optical light, and additionally the pre-peak cooling observed in TDE\,2025aarm and iPTF\,15af, are consistent with predictions from more recent theoretical work on the reprocessing model \citep[e.g.,][]{Giron2026, Mockler26}. \citet{Mockler26} performed time-dependent 1D radiation hydrodynamic simulations that follow a compact X-ray/EUV source reprocessed by an outflow. They find that the optical light curve can lag the bolometric light curve by as much as three weeks. They interpret this lag between the bolometric and optical peaks, and the cooling pre-peak, as a consequence of the time it takes for the reprocessing layer to build up. The peak bolometric luminosity is not observed in TDE\,2025aarm (Fig. \ref{fig:SED_evol}), implying that the optical luminosity lags the bolometric luminosity by at least 20 days.

\subsection{Optical Spectral Evolution}
\begin{figure}
    \centering
    \includegraphics[width=0.8\linewidth]{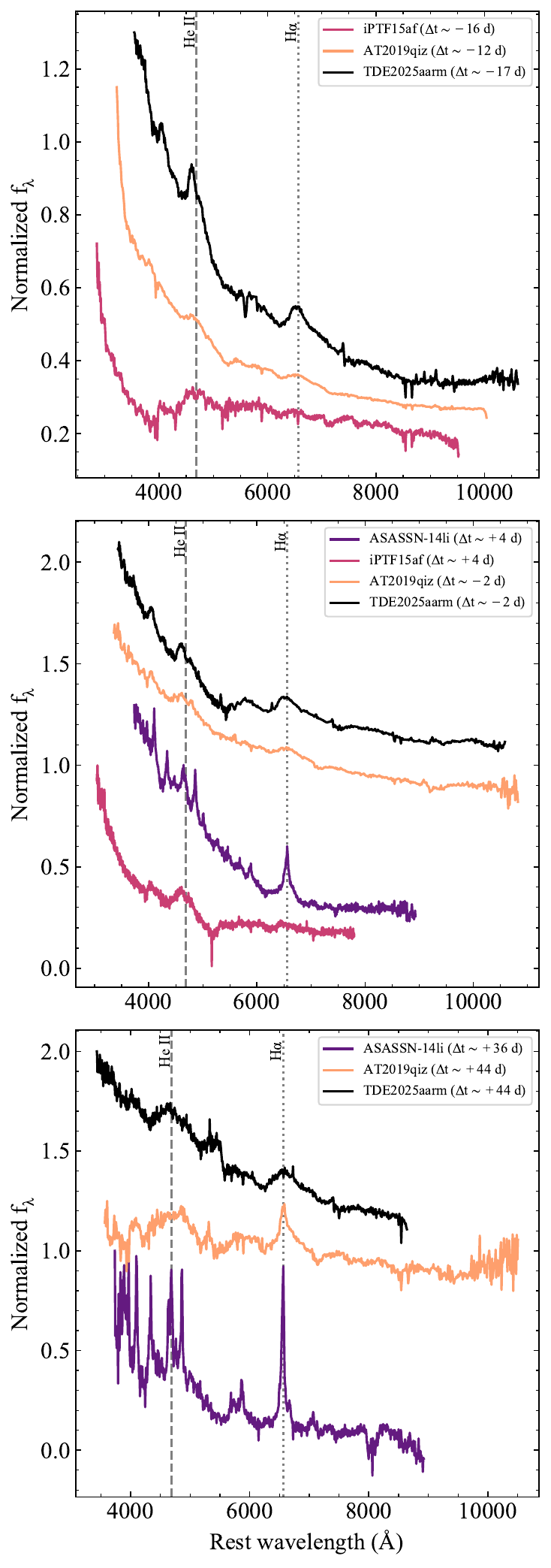}
    \caption{Optical spectral evolution of TDE\,2025aarm compared to selected TDEs \citep{Charalampopoulos22, Hammerstein23, Holoien16_14li} at similar pre-peak, peak, and post-peak epochs. We note that only the spectra of TDE\,2025aarm and iPTF\,15af are host-subtracted. TDE\,2025aarm shows little evolution in the H$\alpha$ line profile, while TDEs such as AT\,2019qiz evolve significantly, becoming narrower at later times.}
    \label{fig:spec_evo}
\end{figure}

\begin{figure}
    \centering
    \includegraphics[width=\linewidth]{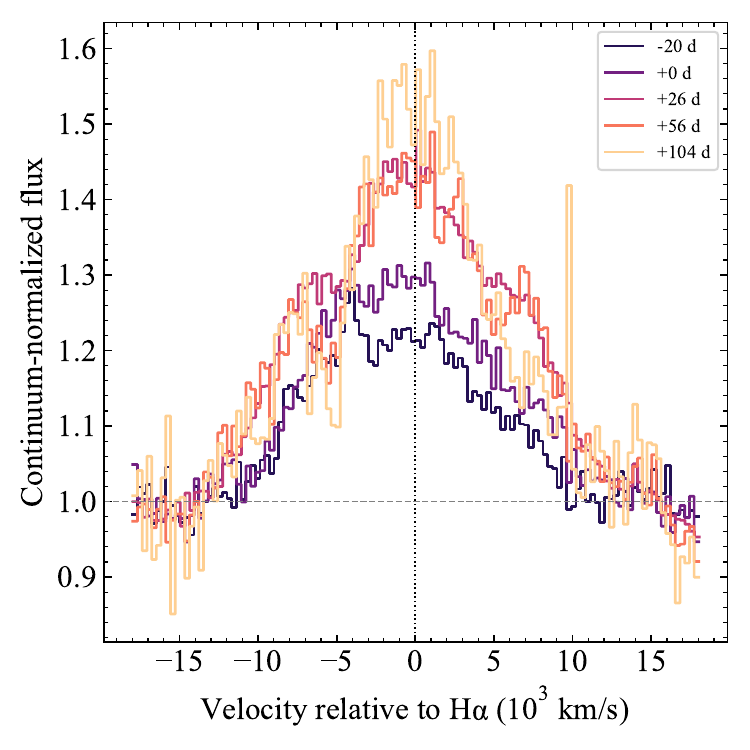}
    \caption{Evolution of the H$\alpha$ line from pre- to post-peak in TDE\,2025aarm. The total line flux evolves significantly but remains broad over the course of the light curve.}
    \label{fig:Halpha_evo}
\end{figure}

In Figure \ref{fig:spec_evo}, we show a comparison of optical spectra at pre-peak, peak, and post-peak epochs for TDE\,2025aarm, iPTF\,15af, AT\,2019qiz, and ASASSN-14li. These are the same objects shown in Figure \ref{fig:spec-comp}, which we use as comparisons to the UV spectrum of TDE\,2025aarm. Spectra of iPTF\,15af, AT\,2019qiz, and ASASSN-14li have been published (respectively) by \citet{Charalampopoulos22}, \citet{Hammerstein23}, and \citet{Holoien16_14li}, and are gathered from WISeREP \citep{yaron12}\footnote{\url{https://www.wiserep.org}}. The evolution of the optical emission-line features in TDE\,2025aarm is slow, with H$\alpha$ and \ion{He}{2} remaining broad and slightly blueshifted from pre- to post-peak. This is shown in Figure \ref{fig:Halpha_evo}, where the H$\alpha$ line remains broad from pre- to post-peak. This is largely consistent with what previous studies of the optical spectral evolution of TDE\,2025aarm have found \citep[e.g.,][]{Simongini_Brutus, Baldini_Brutus, Aamer2026}. \citet{Aamer2026} additionally find that at least 20 days after peak, the H$\alpha$ line profile exhibits deviations from a single Gaussian model which they interpret as the emergence of a disk component. The slow evolution of the broad transient emission lines is in stark contrast to the evolution seen in AT\,2019qiz and ASASSN-14li, whose emission lines gain a narrower peak at later epochs.

The variety in these profiles has been attributed to a combination of viewing-angle and optical-depth effects, where the electron-scattering photosphere is larger in the disk direction than in the polar direction, thus producing broader lines \citep{roth18, Dai2018, leloudas19, nicholl19}. The evolution from broad to narrow profiles may then be explained as a thinning of the photosphere as the debris fallback rate and accretion rate drop post-peak. In this framework, TDEs with the broadest optical lines are those viewed in the disk direction where the photosphere is denser. In the viewing-angle unification picture of \citet{Dai2018}, this scenario should also produce strong outflow signatures and weak or absent X-rays, as observed in TDE\,2025aarm. In this framework, a lack of significant line evolution, as seen in TDE\,2025aarm, would indicate a sustained reprocessing photosphere. At later times, $\sim125$ days after optical peak, the X-rays in TDE\,2025aarm were observed to brighten \citep[Figure \ref{fig:xray-opt};][]{Brutus_Chandra2}. Figure \ref{fig:opticalspec} shows that by 114 days post-peak, the broad optical lines have begun to fade, though remain slightly broad, which may indicate that the reprocessing photosphere has thinned enough to allow some X-rays to emerge.

This lack of evolution of the line profiles could also indicate that there is more mass in the outflow responsible for the reprocessing compared to those events that show significant evolution. In this case, TDEs with persistent broad lines are expected to evolve on longer timescales as a result of the longer photon-diffusion time. \citet{vanVelzen21} found that the rise timescale for iPTF\,15af, which shows little optical spectral evolution in Figure \ref{fig:spec_evo}, was indeed longer than that of AT\,2019qiz. \citet{Blagorodnova2019} also noted the long rise time in iPTF\,15af of $\sim$60 days. If we compare the $t_{\rm 1/2, rise}$ of TDE\,2025aarm to those presented by \citet{Yao2023}, it falls on the higher end of the population, though this might be expected given that the black hole mass is $\sim10^7 M_\odot$. The black hole mass of iPTF\,15af is similar, with $M_{\rm BH} = 10^{6.88} M_\odot$ \citep{Wevers2017}, while \citet{Hammerstein23_IFU} report a lower mass for AT\,2019qiz of $10^{6.23} M_\odot$. Disentangling the different effects of black hole mass, outflow mass and velocity, and viewing angle on the resulting line profiles and evolution will require a more systematic study of line evolution and light curve timescales.

\subsection{Early, Faint X-rays}
\begin{figure}
    \centering
    \includegraphics[width=\linewidth]{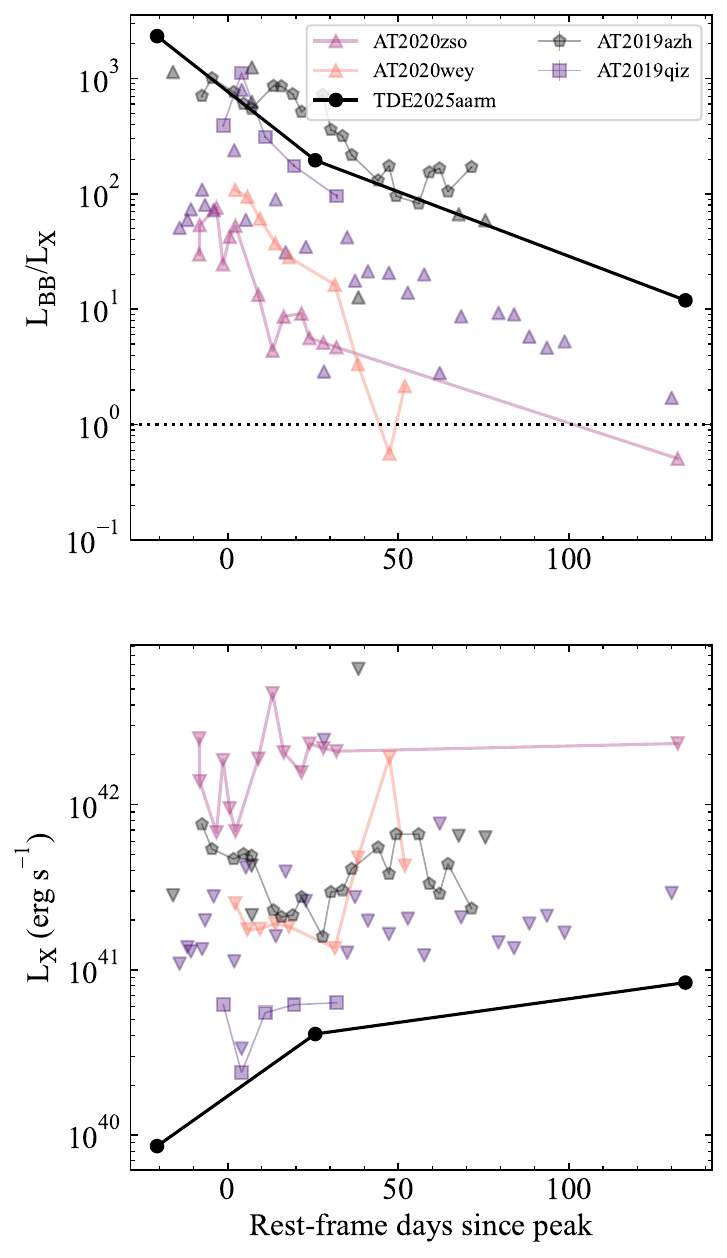}
    \caption{\textit{Top}: Optical/X-ray ratio for TDE\,2025aarm and selected TDEs from \citet{Hammerstein23} with pre-peak Swift XRT observations. TDE\,2025aarm exhibits the largest optical to X-ray ratio at early times by several orders of magnitude, following similar trends toward unity at later times. We note that the blackbody luminosity used here is from fits to the UV/optical photometry. Thus this ratio represents only a conservative estimate of the optical/X-ray ratio. Upward-pointing triangles represent lower limits on the optical/X-ray ratio. AT\,2020zso and AT\,2020wey are only presented here as lower limits, meaning they were not detected in X-rays at any point. \textit{Bottom}: X-ray light curves for the same selected TDEs and TDE\,2025aarm. The Chandra observations of TDE\,2025aarm represent the faintest and earliest detected X-ray luminosity of an optically-selected TDE yet.}
    \label{fig:xray-opt}
\end{figure}

At early times, TDE\,2025aarm exhibits the highest optical/X-ray ratios ever observed, largely due to the detection of one of the faintest X-ray luminosities of an optically-selected TDE yet. In Figure \ref{fig:xray-opt}, we show the evolution of the ratio of the blackbody luminosity to the X-ray luminosity of TDE\,2025aarm, relative to optical light curve peak and the X-ray light curve. We note that this blackbody luminosity is derived from the single blackbody fits to the UV/optical photometry, and thus may be an underestimate of the true bolometric luminosity. We also show TDEs from \citet{Hammerstein23} for which pre-peak X-ray observations with Swift/XRT were made. While limits for AT\,2019azh reach nearly the same level as TDE\,2025aarm 15 days pre-peak, the earliest detected optical/X-ray ratio in AT\,2019azh at 10 days pre-peak is lower by a factor of nearly 3.

Our analysis of the UV spectrum suggests an early launch of a fast moving outflow ($\sim 10,000 \, \rm km \, s^{-1}$). The outer parts of this outflow, if traveling through a dense medium, will result in bright radio emission which is typically associated with relativistic electrons at the shock front (\citet{Brutus_VLA} reported bright radio emission from TDE\,2025aarm). Optical/UV photons from the inner regions can be up-scattered to hard X-rays by these relativistic electrons by the process of inverse-Compton (IC) scattering (as seen, for example, in some core-collapse SNe; \citealt{2006ApJ...641.1029C}), producing an X-ray spectrum that follows a power-law with a hard photon index. This is thus a plausible scenario for producing the radio and X-ray detections in TDE\,2025aarm at early times. \citet{Matsumoto_Brutus} investigated this scenario, finding that both the early radio and X-ray emission may be explained through a narrowly collimated outflow associated with the unbound debris, where the same shock that produces the radio emission would accelerate relativistic electrons that then produce X-rays through either synchrotron or inverse-Compton scattering of UV/optical photons. We use their equation (10) to estimate the total number density of the non-thermal electrons and rescale it to our observed properties:

\begin{align}
    n_{\rm e, \, tot} \simeq & 4 \times 10^4 \, {\rm cm^{-3}} \, \left( \frac{L_{\rm IC}}{8.4 \times 10^{39} \, {\rm erg \, s^{-1}}} \right) \left( \frac{\Omega}{4 \pi} \right)^{-1} \\
    \nonumber & \times \left( \frac{\gamma}{30} \right)^{p-3} \left( \frac{L_{\rm opt}}{1.4 \times 10^{44} \, {\rm erg \, s^{-1}}} \right)^{-1} \\
    \nonumber & \times \left( \frac{v_{\rm sh}}{10^4 \, {\rm km \, s^{-1}}} \right)^{-1} \left( \frac{t}{44 \, {\rm days}} \right)^{-1}.
\end{align}
We use here the optical/UV bolometric luminosity and the X-ray luminosity on 2025 Nov. 7 ($\sim 44$ days after optical discovery) as the luminosity of the seed photons, $L_{\rm opt}$, and of the IC luminosity, $L_{\rm IC}$. We also assume free expansion with a velocity of $\sim 10,000 \, \rm km \, s^{-1}$, a Lorentz factor for the electrons of $\gamma = 30$, and a spherical outflow ($\Omega = 4 \pi$). We find that if the X-ray emission is indeed due to IC scattering of the optical/UV photons, the total number density of the nonthermal electrons is $n_{\rm e, \, tot} \sim 4 \times 10^4 \, \rm cm^{-3}$, which is reasonable compared to the densities inferred from radio-detected TDEs (see e.g., \citealt{Alexander_2020}).

In this scenario where the X-rays are produced via inverse Compton scattering of the UV/optical photons, one might expect an X-ray light curve that declines as the optical light curve declines. This is not seen in TDE\,2025aarm, where post-peak Chandra observations show a rising and softening X-ray light curve \citep{Brutus_Chandra2}. However, if the outflow subsides post-peak and the post-peak optical light curve transitions to a plateau dominated by disk emission \citep[e.g.,][]{vanVelzen2019b}, this could explain the softening and rising of the X-ray light curve at late times.

Alternatively, in the simulations of \citet{Mockler26}, where wind reprocessing is responsible for producing the optical light curve, X-rays from the disk can be observed at very early times, but they fade quickly as the mass outflow builds up around the emitting source. This could be consistent with faint X-rays seen in TDE\,2025aarm, as the outflow is clearly present by the time X-ray observations were taken.

\subsection{Outflows}
In Figure \ref{fig:spec-comp}, we show the UV spectrum of TDE\,2025aarm compared to three other TDEs with observed UV spectra: AT\,2019qiz, iPTF\,15af, and ASASSN-14li. The vast majority of TDEs with UV spectra have shown BAL signatures at various times in their evolution. ASASSN-14li, however, was observed to exhibit strong, broad emission lines \citep[BELs;][]{cenko16}. This BAL versus BEL dichotomy has been seen in Type 1 quasars and explained as an orientation effect. Motivated by this, \citet{Parkinson2020} investigated whether orientation could explain the differences seen in TDE UV spectra. They found that when systems are viewed through the wind ($i=65^\circ$), BALs are observed, and when they are viewed outside of the wind cone ($i=10^\circ$ or $i=75^\circ$), BELs are observed. Given that a large majority of TDEs show evidence for BALs, particularly at earlier times and even pre-peak in TDE\,2025aarm, this high incidence rate of BALs might imply large covering fractions for TDE outflows. Indeed, \citet{Parkinson2020} found that a wide-angle wind model with a covering factor of $f_\Omega = 0.52$ provides a much better match to observations than the typical quasar-inspired model with $f_\Omega = 0.20$.

\subsection{SED Shape} \label{SubSec:SED-discussion}
We show in Figure \ref{fig:continuum-fit} that the FUV--IR SED of TDE\,2025aarm is not well described by a blackbody. Radiation-transport simulations have indeed shown that the SED resulting from reprocessing of X-ray photons from the nascent TDE accretion disk will generally not be well represented by a single-temperature blackbody \citep[e.g.,][]{Roth2016, Parkinson2022, Thomsen2022, Giron2026, Mockler26}. This deviation from a blackbody is difficult to observe without broad wavelength coverage from the FUV to the IR. NUV through optical photometry, which has historically been more accessible for the majority of TDEs, is generally well-fit by a single blackbody \citep[e.g.,][]{vanVelzen21, Hammerstein23}. \citet{Patra2026} obtained an IR spectrum of TDE\,2025abcr, noting that even with a blackbody constrained from only NUV and optical data, there was a significant deviation from a single-temperature blackbody into the IR. Their fitted spectral slope was much shallower than the canonical Rayleigh-Jeans tail. They note that while the data could not definitively determine the origin of the IR excess, one possibility is from free–free emission within a dense, stratified reprocessing layer or outflow as shown by \citet{Roth2016}. Interestingly, our fitted power-law slope for the FUV--infrared SED of TDE\,2025aarm ($f_\lambda \propto \lambda^{-3.08}$) is close to the slope of model continua in \citet{Roth2016}, where $L_\lambda \propto \lambda^{-3}$, though the exact resulting slope is highly dependent on factors such as the density structure, temperature, ionization, opacity, and system geometry.

Following the dense, stratified reprocessing layer framework of \citet{Roth2016} and its luminosity formulation from \citet{Margutti2019} \citep[see also][]{LeBaron2026}, \citet{Patra2026} estimate the density profile of the reprocessing outflow in TDE\,2025abcr from their fitted IR slope ($\nu L_{\nu} \propto \lambda^{-2.13}$), finding a steep profile of $\rho \propto r^{-5.4}$. Since our fitted UV--IR slope for TDE\,2025aarm is comparable ($\nu L_{\nu} \propto \lambda^{-2.08}$), we derive a similarly steep density profile with $\rho \propto r^{-5.1}$. As noted by \citet{Patra2026}, this would imply a wind with a strongly decreasing mass-loss rate with increasing radius.

\citet{Roth2016} found that varying the mass in the wind changes the amount of emission reprocessed to the optical; however, the change in the optical continuum shape is only noticeable once the mass in the wind drops so low that there is hardly any reprocessed emission. Modeling by \citet{Mockler26}\footnote{\citet{Mockler26} only include H, He, and O in their simulations and so the FUV metal absorption lines are absent in their resulting spectra.} is able to produce a reasonable match to the observed SED in TDE\,2025aarm, with a mass in the wind of $\sim0.086$--0.1\,$M_\odot$ and a true bolometric luminosity of $\sim 1.8\times 10^{44}$ erg s$^{-1}$. Combined with the velocities of the BALs in the UV spectrum, this estimated mass would imply a kinetic energy of $\sim10^{50}$ erg, which can be compared to future analyses of the radio-emitting outflow. The value of the bolometric luminosity from their models is much closer to, but slightly higher than, our estimate of the bolometric luminosity from the UV continuum, as opposed to the lower estimate from the single-temperature blackbody fits to the UV/optical photometry. The stark difference is illustrated in Figure \ref{fig:superSED}, where the \citet{Mockler26} model and our blackbody estimated from the UV continuum peak at much shorter wavelengths. While our estimate from the UV continuum is likely closer to the true bolometric luminosity, it is clear that it still misses a nonnegligible fraction of emission in the optical and IR.

This deviation from a blackbody can lead to large underestimation of the true bolometric luminosity of UV/optical emission. In Figure \ref{fig:superSED}, we show that the blackbody estimated from only the NUV and optical photometry underestimates the FUV flux and would likely miss a significant fraction of the EUV flux. This results in an underestimate of the bolometric luminosity by a factor of $\sim9$. This discrepancy has been noted in previous studies of TDE UV spectra \citep[e.g.,][]{Blagorodnova2019, Hung2021} and has been cited as a potential partial resolution of the ``missing energy problem'' \citep[e.g.,][]{Lu2018}, where the observed optical and
UV luminosity in TDEs is much lower than the predicted bolometric luminosity from a black hole accreting at the Eddington limit. If we instead use only the UV data to constrain the source blackbody, which may be closer but still not entirely representative of the true source luminosity \citep[e.g.,][]{Roth2016, Mockler26}, we obtain a hotter and more luminous result that predicts far more EUV flux. If this represents something close to the true continuum level through the UV, it would imply that the majority of the FUV continuum is heavily absorbed. While it is difficult to estimate the true bolometric luminosity and continuum shape without knowledge of the source in the EUV, our results highlight the need for and benefit of observations even into the FUV. Additionally, future observations of a potential dust echo in TDE\,2025aarm would help to place constraints on the total emitted energy and determine whether our estimate for the bolometric luminosity predicted from the UV continuum is a better representation of the true source luminosity.

\section{Summary}
In this work, we present X-ray--IR follow-up observations of the nearby TDE\,2025aarm, including the first-ever UV spectrum of a TDE taken before maximum light. Our main results and conclusions are as follows.
\begin{enumerate}
    \item TDE\,2025aarm exhibits strong evidence for a fast (10,000 km s$^{-1}$) outflow as early as 20 days before optical light-curve peak through blueshifted FUV and NUV BALs, providing further confirmation that TDEs are capable of launching fast outflows at early times.
    \item We fit different portions of the FUV--optical continuum with a blackbody and find that fitting a single-temperature blackbody to the NUV/optical photometry alone may underestimate the true bolometric luminosity by a factor of $\sim$9. This emphasizes the need for FUV observations of TDEs that can probe the SED closer to its peak and capture the true energetics of these events.
    \item The slope of the FUV--IR SED of TDE\,2025aarm is much shallower than the Rayleigh-Jeans tail, with $f_\lambda \propto \lambda^{-3.08}$. We interpret this as a result of reprocessing through the outflow \citep[e.g.,][]{Roth2016}, which can produce spectral slopes shallower than a single-temperature blackbody with $L_\lambda \propto \lambda^{-3}$.
    \item We find evidence for two BALs that have not yet been identified in a TDE at $\lambda_{\rm obs}\approx1587$ \AA~and $\lambda_{\rm obs}\approx1670$ \AA. The former line may be associated with \ion{He}{2} (strictly from velocity arguments), or it may have contributions from \ion{Fe}{4}. The latter line may be associated with \ion{N}{4}.
    \item We find that the optical light curve peaks at later times relative to the bolometric light curve. Additionally, the blackbody temperature cools significantly pre-peak before stabilizing post-peak. This is consistent with predictions from wind-reprocessing modeling \citep[e.g.,][]{Mockler26}, where the lag between the bolometric light curve and optical light curve, as well as the pre-peak cooling, are a result of the time it takes to build the reprocessing layer.
    \item We find that the pre-peak weak X-ray emission can be reasonably explained by inverse-Compton scattering of the UV/optical photons. The post-peak X-ray brightening/softening is not necessarily inconsistent with this picture if the outflow responsible for the early-time optical emission subsides post-peak and the late-time optical light curve is dominated by accretion disk emission.
\end{enumerate}

Further monitoring of TDE\,2025aarm will probe the evolution of the outflow observed pre-peak, which may help to constrain models for the origin of the UV and optical emission. Our observations of TDE\,2025aarm have demonstrated that FUV coverage is essential to probing the full energetics of TDEs. A larger sample of TDEs with UV spectroscopic coverage is needed to determine whether the early, fast outflow signatures we observe pre-peak in TDE\,2025aarm and the deviation from a blackbody across the wider SED are indeed ubiquitous. Future TDE searches with observatories such as the NSF-DOE Vera C. Rubin Observatory will provide pre-peak $u$ band coverage for nearby TDEs, while its exceptional depth will allow for rest-frame UV coverage of new high redshift TDEs. In the coming years, missions such as the Ultraviolet Explorer \citep[UVEX;][]{UVEX_kulkarni} will allow for systematic follow-up UV spectroscopy of TDEs, enabling studies of their UV spectroscopic evolution on a population scale.

\begin{acknowledgements}

The authors would like to thank Nidia Morrell for help with the reduction of Magellan data. The authors would also like to thank Dan Kasen for helpful guidance regarding the UV line identifications.

Based in part on observations obtained with the NASA/ESA Hubble Space Telescope, retrieved from the Mikulski Archive for Space Telescopes (MAST) at the Space Telescope Science Institute (STScI). STScI is operated by the Association of Universities for Research in Astronomy, Inc. under NASA contract NAS 5-26555. Support for Program number GO-18049 was provided through a grant from the STScI under NASA contract NAS5-26555.
E.H. and R.C. acknowledge support from NASA ADAP Program 80NSSC24K1492 through a subaward from UC Santa Cruz.

Based in part on data obtained with the Samuel Oschin Telescope 48-inch and the 60-inch Telescope at the Palomar Observatory as part of the Zwicky Transient Facility. ZTF is supported by the National Science Foundation under Award \#2407588 and a partnership including Caltech, USA; Caltech/IPAC, USA; University of Maryland, USA; University of California, Berkeley, USA; Cornell University, USA; Drexel University, USA; University of North Carolina at Chapel Hill, USA; Institute of Science and Technology, Austria; National Central University, Taiwan; and the German Center for Astrophysics (DZA), Germany. Operations are conducted by Caltech's Optical Observatory (COO), Caltech/IPAC, and the University of Washington at Seattle, USA. 
The Gordon and Betty Moore Foundation, through both the Data-Driven Investigator Program and a dedicated grant, provided critical funding for SkyPortal.
The ZTF forced-photometry service was funded under Heising-Simons Foundation grant 12540303 (PI M. Graham).
SED Machine is based upon work supported by the National Science Foundation under Grant No. 1106171.

A major upgrade of the Kast spectrograph on the Shane 3\,m telescope at Lick Observatory, led by Brad Holden, was made possible through gifts from the Heising-Simons Foundation, William and Marina Kast, and the University of California Observatories. Research at Lick Observatory is partially supported by a generous gift from Google.

Some of the data presented herein were obtained at Keck Observatory, which is a private 501(c)3 nonprofit organization operated as a scientific partnership among the California Institute of Technology, the University of California, and the National Aeronautics and Space Administration. The Observatory was made possible by the generous financial support of the W. M. Keck Foundation. 
The authors wish to recognize and acknowledge the very significant cultural role and reverence that the summit of Maunakea has always had within the Native Hawaiian community. We are most fortunate to have the opportunity to conduct observations from this mountain.

The scientific results reported in this article are based in part on observations made by the Chandra X-ray Observatory. The authors thank the Chandra PI Pat Slane and the entire Chandra team for enabling these rapid observations. 

A.V.F.'s research group at U.C. Berkeley acknowledges financial assistance from Gary and Cynthia Bengier, Clark and Sharon Winslow, Alan Eustace and Kathy Kwan (W.Z. is a Bengier-Winslow-Eustace Specialist in Astronomy), Timothy and Melissa Draper, Briggs and Kathleen Wood, Ellyn and Alan Seelenfreund (T.G.B. is a Draper-Wood-Seelenfreund Specialist in Astronomy), and numerous other donors.

\end{acknowledgements}

\facilities{HST (STIS), Swift (UVOT), Keck:I (LRIS), Keck:II (NIRES), Shane (Kast), Magellan:Baade (IMACS, MagE), SOAR (GHTS), CXO (ACIS-S), ATLAS, NuSTAR}

\software{HEASoft \citep{heasoft2014}, 
          CIAO \citep{Fruscione2006}, 
          Sherpa \citep{Freeman2001,Doe2007,Siemiginowska2024}, 
          PypeIt \citep{Prochaska2020}, 
          Prospector \citep{Johnson2021}, 
          Spextool \citep{Cushing2004}, 
          xtellcorr \citep{Vacca2003}, 
          emcee \citep{ForemanMackey2013}, 
          nupipeline \citep{Harrison2013}, 
          FSPS \citep{Conroy2009,Conroy2010}, 
          scipy \citep{Virtanen2020}, 
          astropy \citep{AstropyCollaboration2013,AstropyCollaboration2018,AstropyCollaboration2022}, 
          matplotlib \citep{Hunter2007},
          skyportal \citep{vanderWalt2019, skyportal_coughlin}}

\begin{appendix}
\section{SEDM Spectra}
\begin{figure}
    \centering
    \includegraphics[width=0.75\linewidth]{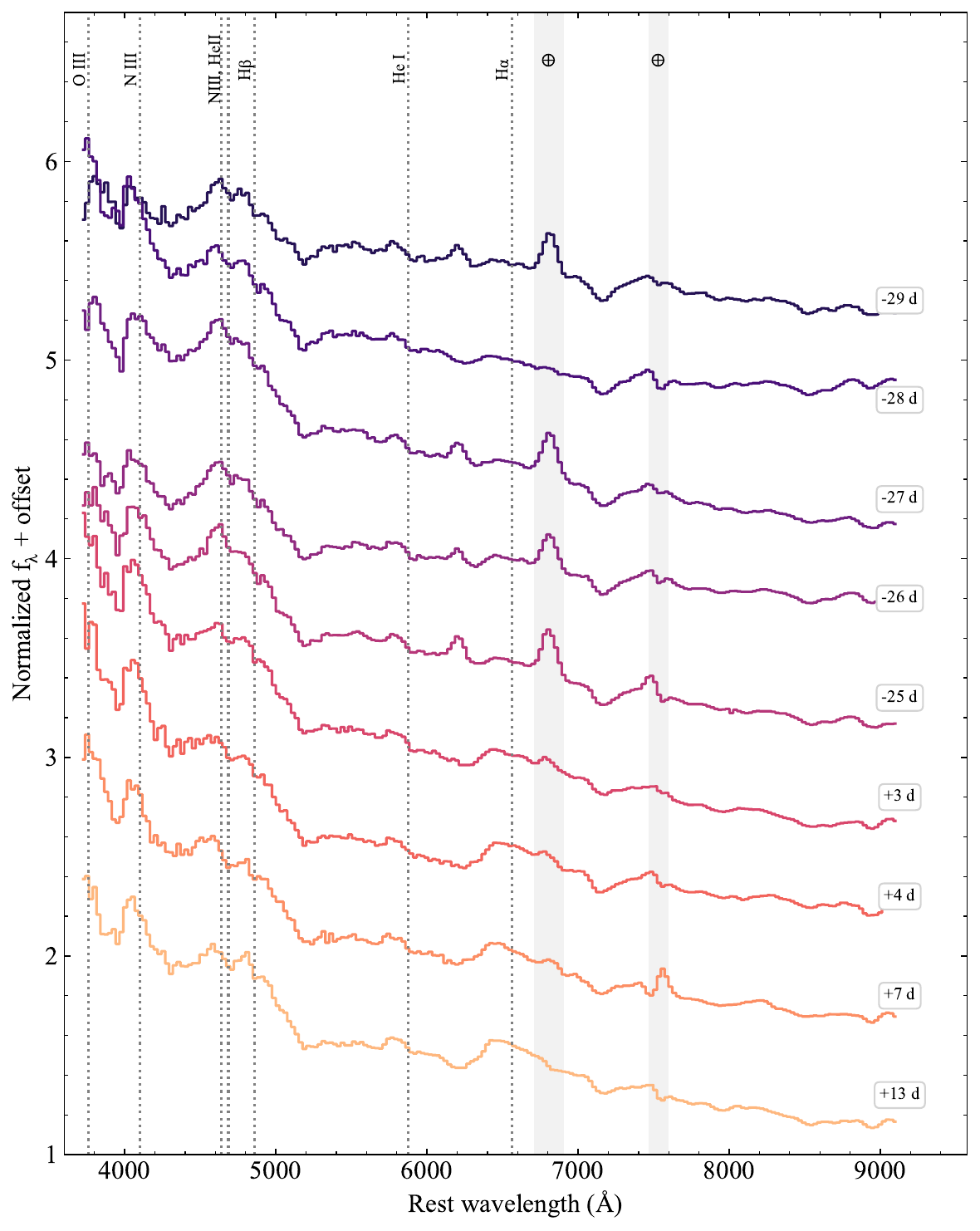}
    \caption{SEDM spectra of TDE\,2025aarm. We mark common TDE broad emission lines and the regions of telluric contamination.}
    \label{fig:sedm}
\end{figure}

We list in Table \ref{tab:SEDM} the spectra taken by SEDM, which are made available as part of the data for this paper but are not included in the analysis due to their low resolution. The spectra are shown in Figure \ref{fig:sedm}.

\begin{deluxetable}{cc}
\label{tab:SEDM}
\tablecaption{SEDM Spectra of TDE\,2025aarm}
\tablehead{
\colhead{Date} & \colhead{$\Delta t$ (days)}}
\startdata
2025-10-29 & $-29$\\
2025-10-30 & $-28$\\
2025-10-31 & $-27$\\
2025-11-01 & $-26$\\
2025-11-02 & $-25$\\
2025-11-30 & +3\\
2025-12-01 & +4\\
2025-12-04 & +7\\
2025-12-10 & +13
\enddata
\end{deluxetable}

\end{appendix}

\bibliography{main, bibliography}{}
\bibliographystyle{aasjournalv7}

\end{document}